\documentclass{aa}
\usepackage{textcomp}
\usepackage{graphicx} %
\usepackage{amsmath} %
\usepackage{amssymb} %
\usepackage{bm} %
\usepackage{upgreek} %
\usepackage{IEEEtrantools} %
\usepackage{multirow}
\usepackage[dvipsnames]{xcolor}
\usepackage[colorlinks=true,allcolors=blue,urlcolor=blue]{hyperref}
\usepackage{txfonts}
\usepackage[normalem]{ulem} 

\newcommand{\sect}[1]{\text{Section~\ref{#1}}}
\newcommand{\fig}[1]{\text{Figure~\ref{#1}}}
\newcommand{\tab}[1]{\text{Table~\ref{#1}}}

\newcommand{\multitd}{\texttt{Multi3D}}
\newcommand{\balder}{\texttt{Balder}}
\newcommand{\scate}{\texttt{Scate}}

\newcommand{\marcs}{\texttt{MARCS}}
\newcommand{\atlas}{\texttt{ATLAS9}}
\newcommand{\stagger}{\texttt{Stagger}}

\newcommand{\cles}{\texttt{CLES}}
\newcommand{\genec}{\texttt{GENEC}}

\newcommand{\lgeps}[1]{A(\mathrm{#1})}
\newcommand{\nm}{\mathrm{nm}}
\newcommand{\dex}{\mathrm{dex}}
\newcommand{\kms}{\mathrm{km\,s^{-1}}}
\newcommand{\vsini}{\mathrm{\varv\sin{\iota}}}

\begin{document} 

\title{The solar beryllium abundance revisited with 3D non-LTE
models\thanks{Table 2 is available in electronic form at the CDS
via anonymous ftp to
\url{cdsarc.u-strasbg.fr (130.79.128.5)} or
via~\url{https://cdsarc.cds.unistra.fr/viz-bin/qcat?J/A+A/}.}}
\author{A.~M.~Amarsi\inst{\ref{uu1}}
\and
D.~Ogneva\inst{\ref{uu1}}
\and
G.~Buldgen\inst{\ref{liege1}}
\and
N.~Grevesse\inst{\ref{liege1},\ref{liege2}}
\and
Y.~Zhou\inst{\ref{chch}}
\and
P.~S.~Barklem\inst{\ref{uu1}}}

\institute{\label{uu1}Theoretical Astrophysics, 
Department of Physics and Astronomy,
Uppsala University, Box 516, SE-751 20 Uppsala, Sweden\\
\email{anish.amarsi@physics.uu.se}
\and
\label{liege1}Space sciences, Technologies and Astrophysics Research (STAR)
Institute, 
Universit\'e de Li\`ege, All\'ee du 6 ao\^ut, 17, B5C, 
B-4000 Li\`ege, Belgium
\and
\label{liege2}Centre Spatial de Li\`ege, Universit\'e de Li\'ege,
avenue Pr\'e Aily, 
B-4031 Angleur-Li\`ege, Belgium
\and
\label{chch}School of Physical and Chemical Sciences ---
Te Kura Mat\=u, University of Canterbury, Private Bag 4800,
Christchurch 8140, Aotearoa, New Zealand}

\abstract{The present-day abundance of beryllium
in the solar atmosphere provides clues about mixing
mechanisms within stellar interiors.
However, abundance determinations based on
the \ion{Be}{II} $313.107\,\nm$ line
are prone to systematic errors due to imperfect model spectra.
These errors arise from missing continuous opacity in the UV,
a significant unidentified blend at $313.102\,\nm$, 
departures from local thermodynamic
equilibrium (LTE), and microturbulence and
macroturbulence fudge parameters
associated with one-dimensional
(1D) hydrostatic model atmospheres.
Although these factors have been discussed 
in the literature, 
no study has yet accounted for all of them simultaneously.
To address this, we present 3D non-LTE calculations for neutral and
ionised beryllium in the Sun.
We used these models to derive the present-day
solar beryllium abundance, calibrating the missing
opacity on high resolution solar irradiance data
and the unidentified blend on the centre-to-limb variation.
We find a surface abundance of $1.21\pm0.05\,\dex$,
which is significantly lower than the value of $1.38\,\dex$
that has been commonly adopted since 2004.
Taking the initial abundance via CI chondrites, our result implies
that beryllium has been depleted from the surface
by an extra $0.11\pm0.06\,\dex$, or $22\pm11\%$,
on top of any effects of atomic diffusion. This
is in tension with standard solar models, which predict negligible depletion,
as well as with contemporary solar models that have
extra mixing calibrated on the abundances of helium
and lithium, which predict excessive depletion.
These discrepancies highlight the need for further improvements to the
physics in solar and stellar models.}


\date{Received 2 August 2024 / Accepted 22 August 2024}
\maketitle

\section{Introduction}
\label{introduction}

Beryllium is an interesting tracer of the structure and 
evolution of the Sun and other late-type stars.
$\mathrm{^{9}Be}$ is destroyed at temperatures of around 
$3.5\times10^{6}\,\mathrm{K}$,
that is, at temperatures slightly greater than that
of $\mathrm{^{7}Li}$ 
($2.5\times10^{6}\,\mathrm{K}$; e.g.~\citealt{2015ApJ...811...99L}).
Given that the temperature at the base of the solar convective
zone ranges from around $2.45\times10^{6}\,\mathrm{K}$
to $2.20\times10^{6}\,\mathrm{K}$ from birth 
until today \citep[e.g.][]{2021LRSP...18....2C},
the depletion of beryllium in the solar atmosphere,
and the relative depletion of beryllium compared to lithium,
can help constrain the mixing of material from
the convective envelope into the radiative interior 
\citep[e.g.][]{2022NatAs...6..788E,2023A&A...669L...9B}.

The total cumulative depletion of beryllium in the Sun is determined by
the difference between the
initial abundance and the current surface abundance.  The initial
abundance is typically assumed to be well-represented by CI chondrites
\citep[e.g.][]{1956RvMP...28...53S,1989GeCoA..53..197A,2009LanB...4B..712L}.
Although there are hints
of trends in CI chondrite compositions with condensation temperature
\citep{1997MNRAS.285..403G,2010MNRAS.407..314G,2021A&A...653A.141A};
the effect of this is
minimised by using an abundant species of similar condensation temperature to
beryllium ($T_{\mathrm{c}}=1551\,\mathrm{K}$; \citealt{2019AmMin.104..844W}) to
convert from the meteoritic scale to the solar scale.\footnote{The
protosolar abundance may be slightly higher because this
conversion erases the effects of atomic diffusion
 (microscopic thermal diffusion, gravitational settling, and
radiative acceleration), given that
they are predicted to be of a similar magnitude for all elements heavier than
helium (see Section 5 of \citealt{2021A&A...653A.141A}).} 
Using silicon
($T_{\mathrm{c}}=1314\,\mathrm{K}$),
magnesium ($T_{\mathrm{c}}=1343\,\mathrm{K}$),
or iron ($T_{\mathrm{c}}=1338\,\mathrm{K}$),
with $\lgeps{Si}=7.51\pm0.03$, $\lgeps{Mg}=7.55\pm0.03$,
or $\lgeps{Fe}=7.46\pm0.04$
\citep{2021A&A...653A.141A}, the CI chondrite abundances
from \citet{2021SSRv..217...44L} are
$\lgeps{Be}_{\text{init}}=1.31\pm0.05$, $1.34\pm0.05$, and 
$1.32\pm0.05$, respectively, for the initial solar
composition.\footnote{$\lgeps{X}\equiv\log_{10}N_{\mathrm{X}}/N_{\mathrm{H}}+12$.} Taking the mean, we arrive at 
$\lgeps{Be}_{\text{init}}=1.32\pm0.04$.

The current surface abundance of beryllium must be
determined from analysing the solar spectrum.
The \ion{Be}{II} $313\,\nm$ resonance doublet
is the most useful feature in the spectra of late-type
stars \citep[e.g.][]{2023ApJ...943...40B,2023ExA....55...95S}. 
In the Sun, the stronger component at $313.042\,\nm$ (in air)
is heavily affected by blends.
For the weaker component at $313.107\,\nm$, the focus
of this study, precise and accurate determinations
of the solar beryllium abundance is hindered by four uncertainties
of potentially comparable importance:
a) missing continuous opacity;
b) a significant blend at around $313.102\,\nm$; 
c) errors caused by the assumption
of local thermodynamic equilibrium (LTE); and d) errors caused by
the use of one-dimensional (1D) model atmospheres.

So far, there have not been any
studies of the solar beryllium abundance
to address all four uncertainties simultaneously.
A notable work is that of \citet{1975A&A....42...37C}.
They accounted for missing continuous opacity
by considering the continuous centre-to-limb 
variation, and they accounted for the blend
by considering the shape of the
\ion{Be}{II} $313.107\,\nm$ line
at disc-centre ($\mu=1.0$) and at the limb ($\mu=0.2$).
In this way they found that the blend
constitutes almost half of the equivalent width
at the limb (see their Figure 6).
They carried out non-LTE calculations with 
a small nine-level
model atom for neutral and ionised beryllium
using a revised version of the 1D semi-empirical
model atmosphere of \citet{1974SoPh...39...19H}. 
They ultimately arrived at $\lgeps{Be}=1.15\pm0.20$.
The mean value
implies a depletion of around $30\%$, although
the large error bar makes it consistent with no depletion
to one standard error.

The later work by \citet{1998Natur.392..791B}
took a different approach
to calibrating for any missing continuous opacity:
namely, they considered 
A-X electronic transitions of OH close to the \ion{Be}{II}
doublet, and demanded that they yield
the same oxygen abundance as vibrational-rotational
transitions in the infrared.
Their subsequent analysis was based solely
on disc-integrated flux (rather than the centre-to-limb variation as in 
\citealt{1975A&A....42...37C}) and
they did not discuss the impact of the blend.
They also did not discuss departures from LTE,
and they used the 
1D semi-empirical atmosphere of \citet{1974SoPh...39...19H}. 
In all they obtained $\lgeps{Be}=1.40\pm0.09$,
consistent with no depletion of beryllium in the Sun.

\citet{2004A&A...417..769A} presented a reanalysis
of \citet{1998Natur.392..791B}.
The primary development was the use of a 
3D hydrodynamical model atmosphere,
both for calibrating the missing continuous opacity via
OH lines, and for determining the beryllium abundances.
\citet{2004A&A...417..769A} arrived at $\lgeps{Be}=1.38\pm0.09$:
again, consistent with no depletion of beryllium
in the solar convective envelope.
This value is also given in the compilation of
\citet{2021A&A...653A.141A}, although 
it is cautioned that this value is likely to be overestimated
owing to the limitations in the original studies
of \citet{1998Natur.392..791B} and
\citet{2004A&A...417..769A}:
in particular, they neglected
the impact of the blend as well as departures from LTE.

In a more recent study, \citet{2018ApJ...865....8C}
analyse the disc-integrated flux from the Sun
observed from the asteroid Vesta,
as well as spectra from the red giants
Arcturus and Pollux.
They updated a line list around the \ion{Be}{II} $313\,\nm$ lines
from \citet{2012ApJ...746...47D}
and \citet{2011PASJ...63..697T}
(the latter originally developed by
\citealt{1997ApJ...480..784P}) to
fit these three spectra using a theoretical \marcs{} 1D model atmosphere 
\citep{2008A&A...486..951G}, and 1D LTE spectrum synthesis.
In doing so they argue that the unidentified blend is probably 
due to an ionised species, and is of low excitation potential --- 
just as \citet{1975A&A....42...37C} argued on the basis
of the solar centre-to-limb variation.
Using their calibrated line list, they derived a solar abundance
of $\lgeps{Be}=1.30$,
consistent with no depletion of beryllium.
Nonetheless they cautioned
of large systematic uncertainties, largely stemming from the 
calibration of the missing continuous opacity via an OH line.

A very recent work on this topic is that of
\citet{2022A&A...657L..11K}. Their analysis, like that of 
\citet{1998Natur.392..791B} and
\citet{2004A&A...417..769A}, was based on the disc-integrated flux.
The missing continuous  opacity was calibrated by comparing against
the observed low-resolution absolute disc-integrated flux in the near-UV region.
Rather than adding hypothetical lines, the authors start with a
predicted line list from the VALD database
\citep[e.g.][]{2008JPhCS.130a2011H},
and modified the oscillator strengths and wavelengths
of blending lines mainly on the basis
of literature recommendations (which in turn are based
on 1D LTE analyses of the Sun and other stars).
In particular, the unidentified blend
was taken to be the known \ion{Mn}{I} $313.104\,\nm$ line,
with its oscillator strength and wavelength adjusted
to the values given by \citet{2005MSAIS...8..206A}.
With this line list, the authors presented a non-LTE analysis
of neutral and ionised beryllium based on an extensive
model atom with up-to-date collisional data.
They arrived at $\lgeps{Be}=1.32\pm0.05$
and $\lgeps{Be}=1.35\pm0.05$, using a theoretical \atlas{}
1D model atmosphere \citep{2003IAUS..210P.A20C}
and the 1D semi-empirical
model atmosphere of \citet{1974SoPh...39...19H}, respectively.
Again, there is no indication of significant depletion of 
beryllium in the solar atmosphere.

Here, we revisit the solar beryllium abundance.
We attempted to take into account
all of the problems mentioned above in a single study:
namely, the missing continuous opacity, and the $313.102\,\nm$ blend,
together with 3D non-LTE effects.
We first present an overview of the 3D non-LTE 
calculations in \sect{method},
before presenting our analysis of the solar beryllium
abundance in \sect{results}. 
We ultimately found evidence of some beryllium depletion in the solar
envelope; we discuss the implications of this result 
on the structure and evolution of the Sun in
\sect{discussion}, and then summarise our findings in
\sect{conclusion}.

\section{Models}
\label{method}

\begin{figure*}
    \begin{center}
        \includegraphics[scale=0.325]{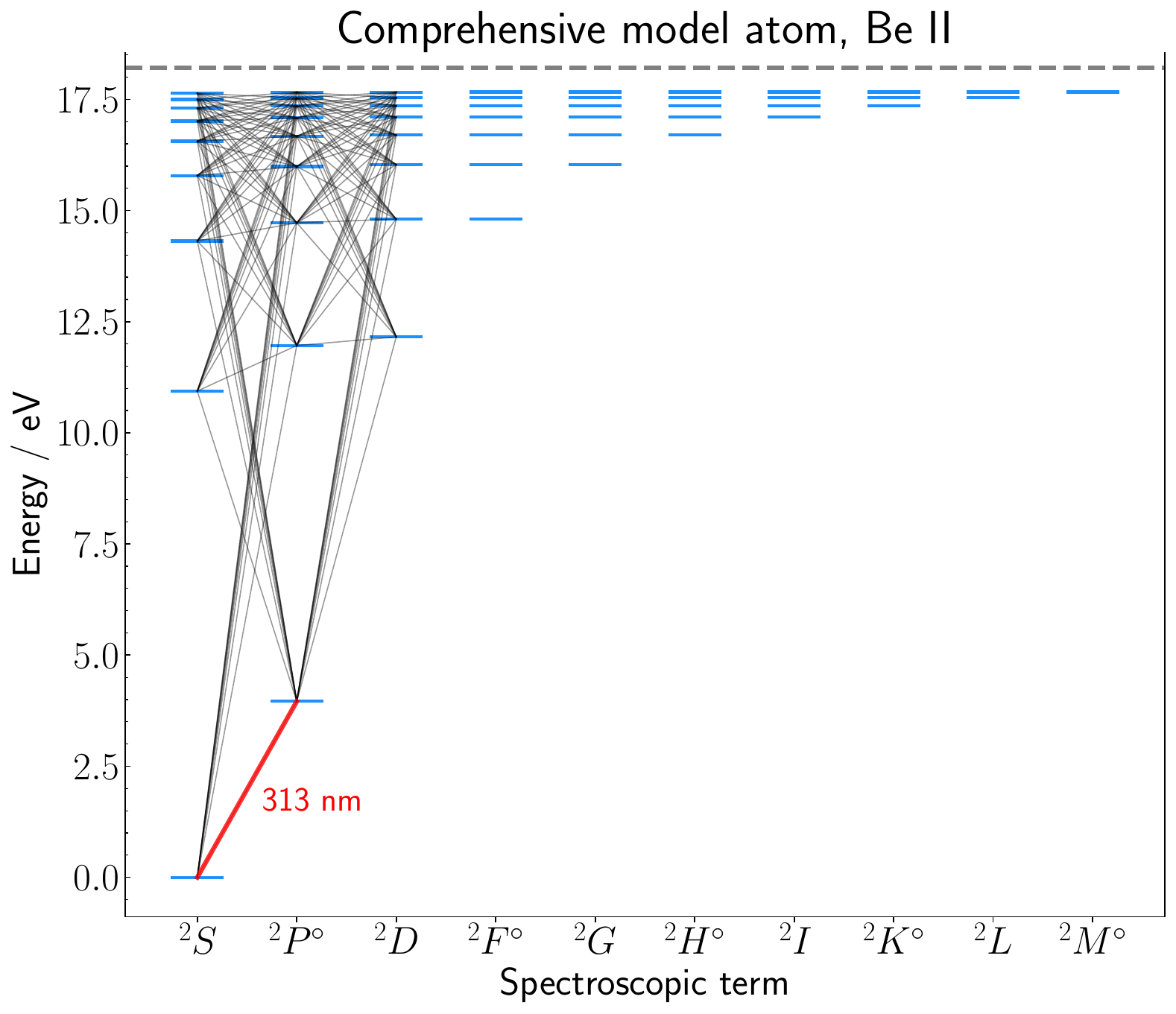}
        \includegraphics[scale=0.325]{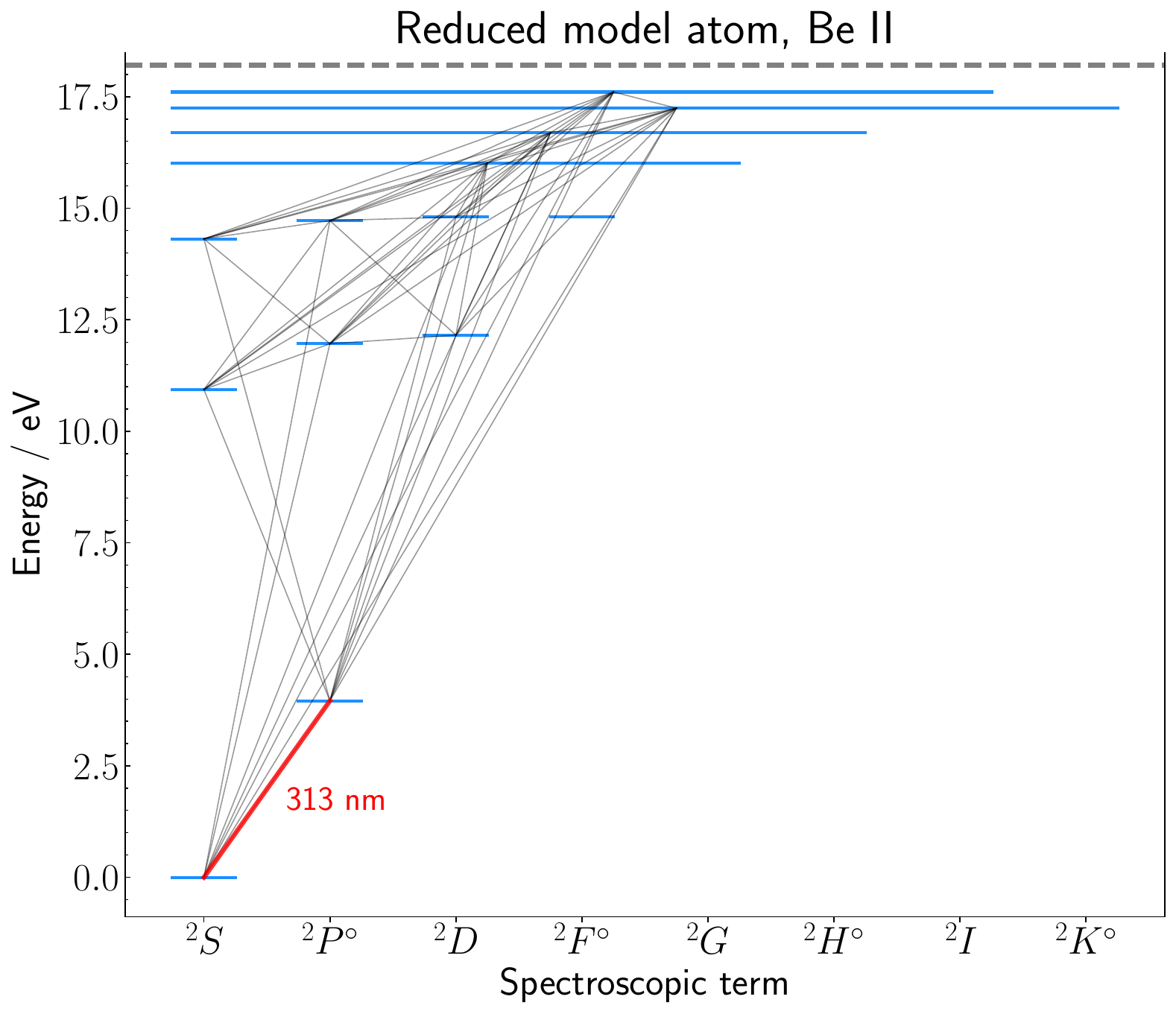}
        \includegraphics[scale=0.325]{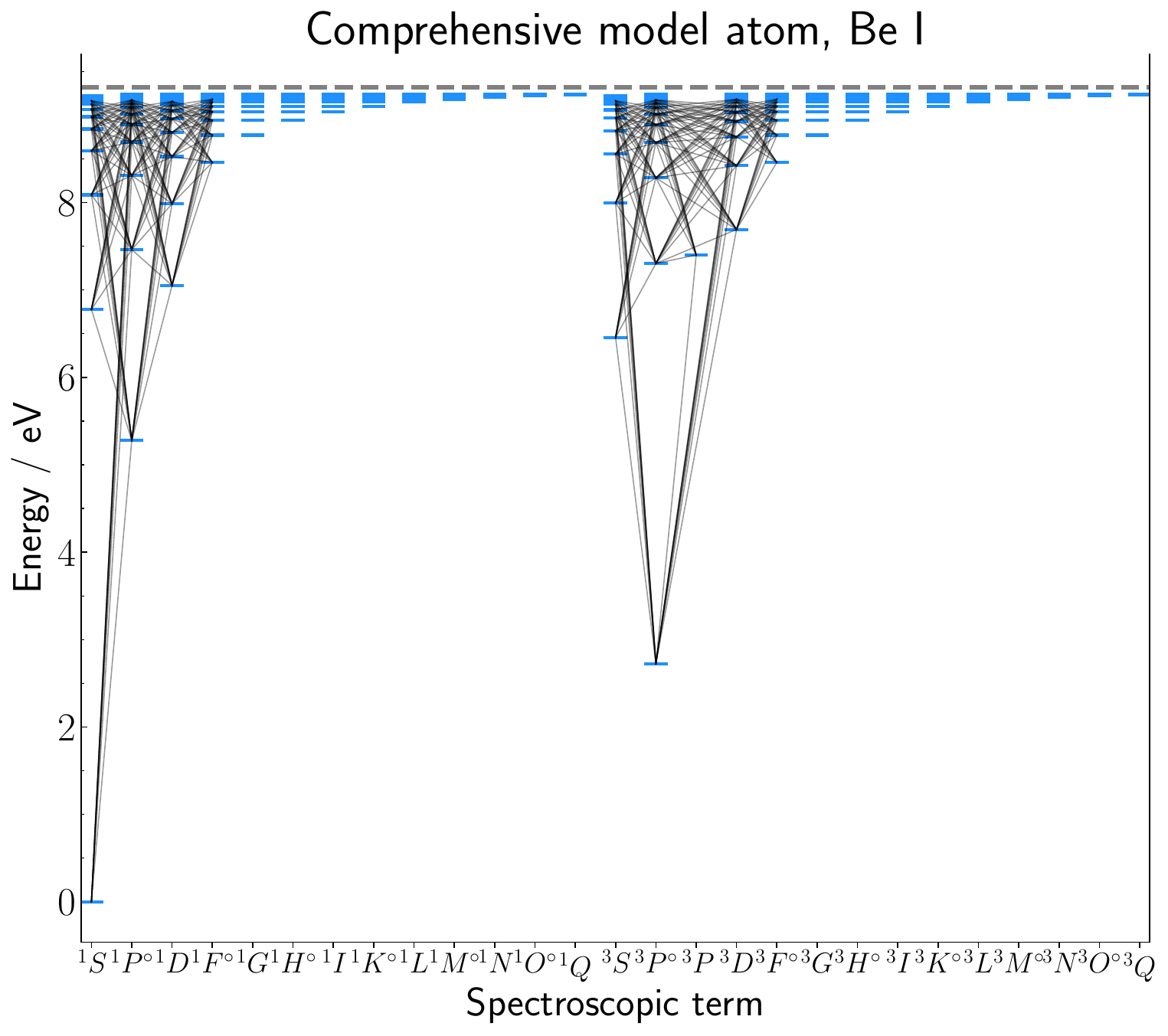}
        \includegraphics[scale=0.325]{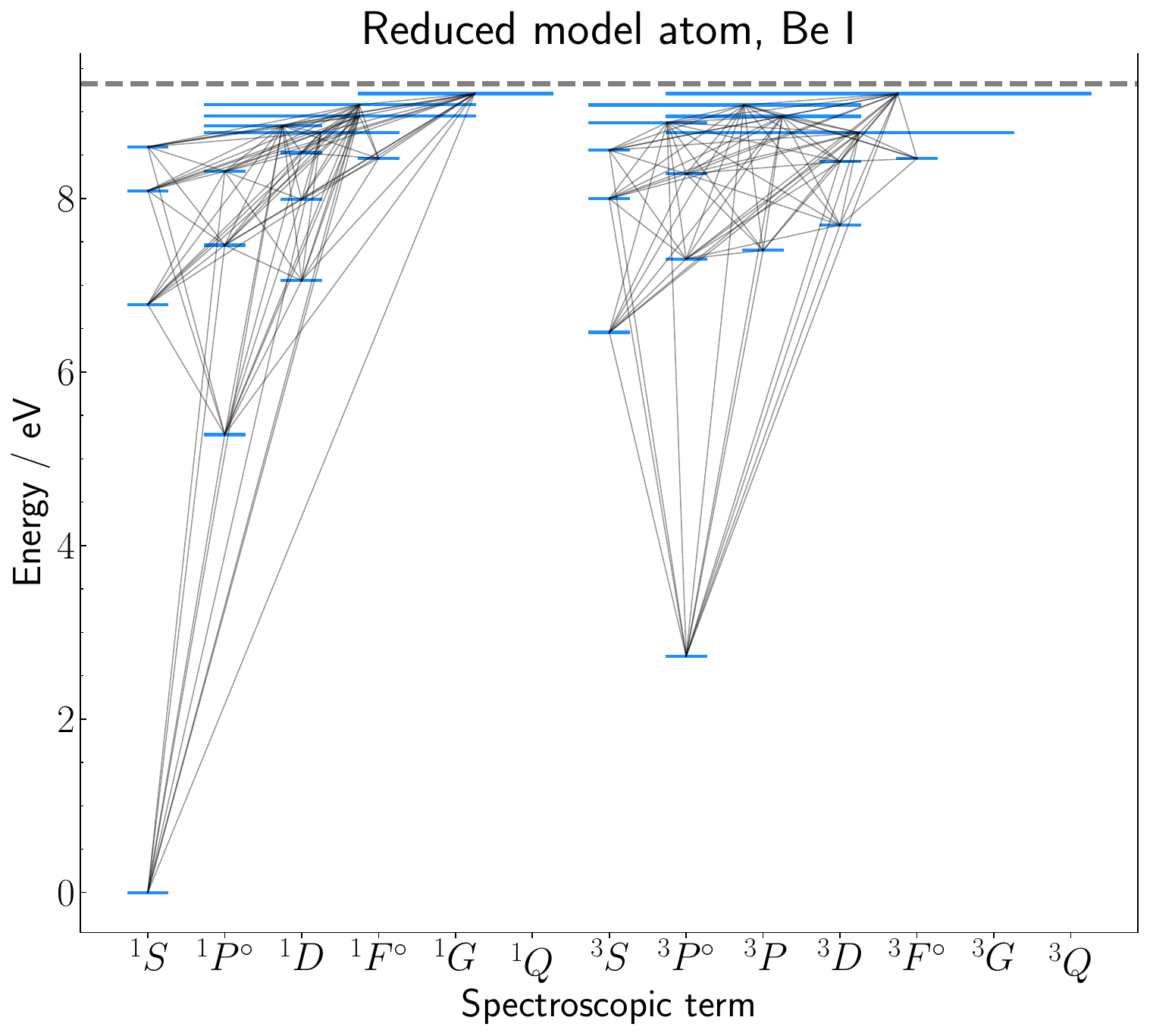}
        \caption{Grotrian diagrams of singly ionised (upper panels) 
        and neutral (lower panels)
        beryllium as represented in the 
        comprehensive (left panels) and reduced (right panels)
        model atoms used in this work.}
        \label{fig:atom}
    \end{center}
\end{figure*}

\begin{figure*}
    \begin{center}
        \includegraphics[scale=0.325]{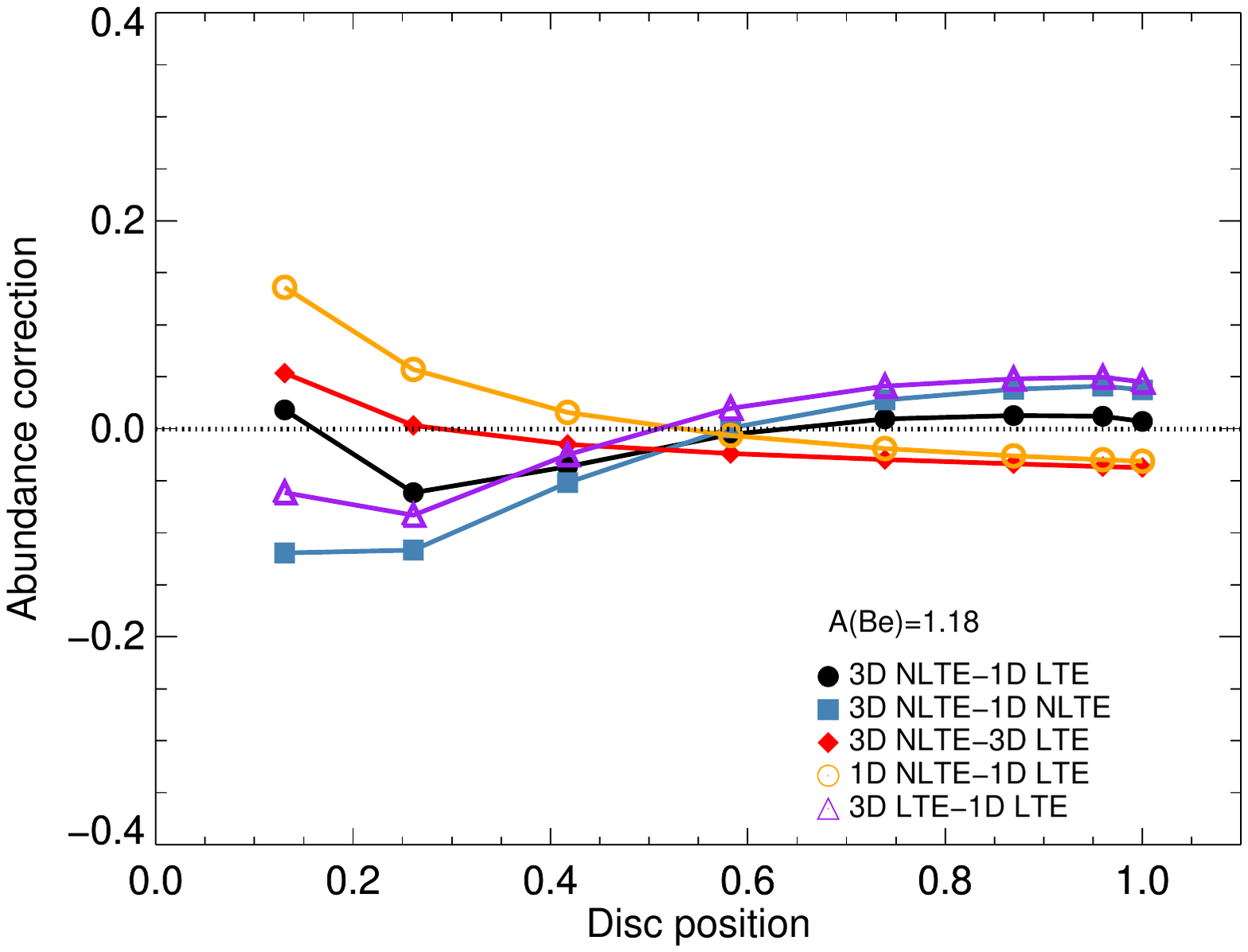}
        \includegraphics[scale=0.325]{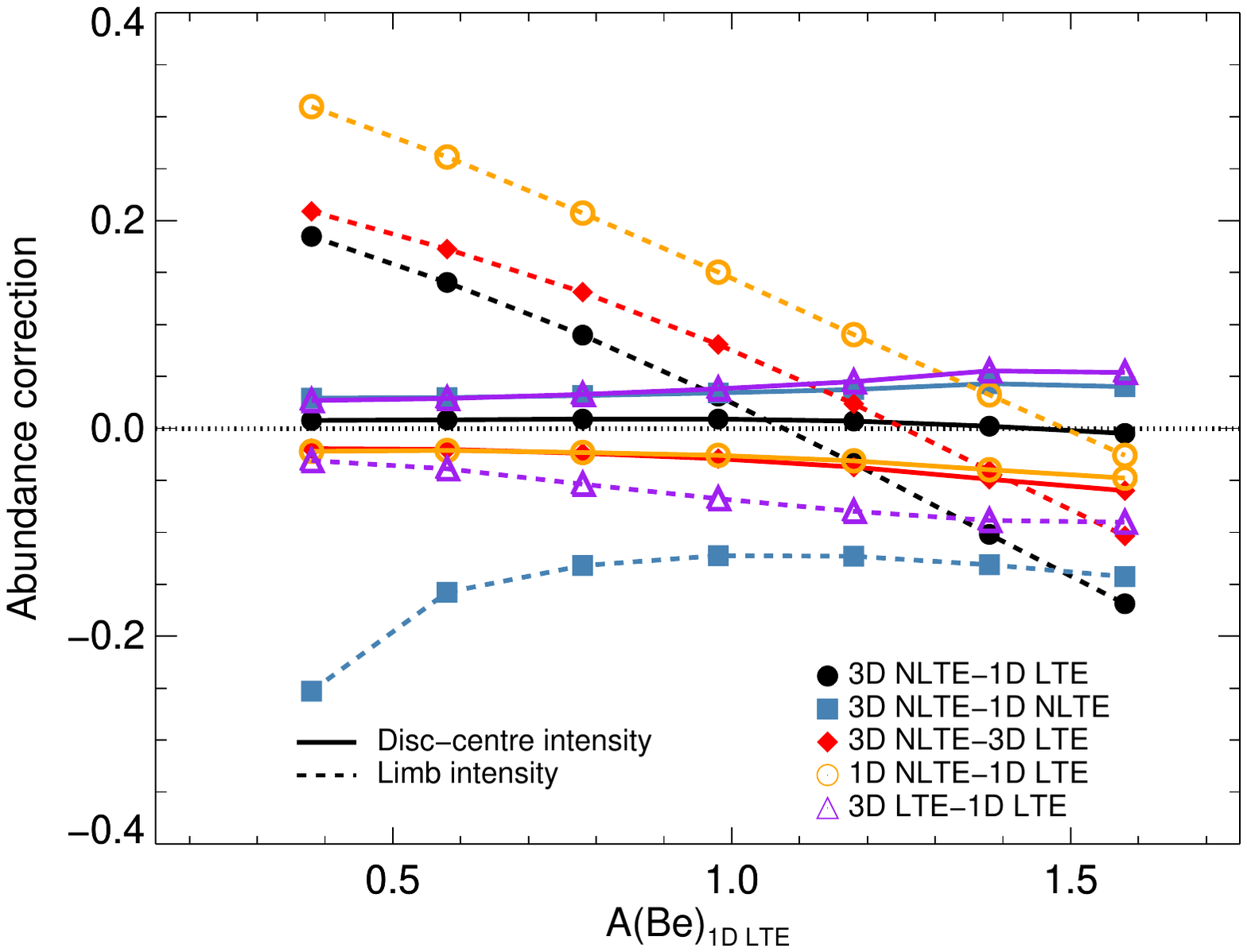}
        \caption{Abundance corrections for the \ion{Be}{II} $313.107\,\nm$,
        based on equivalent widths
        as a function of $\mu$ at a reference abundance of
        $\lgeps{Be}=1.18$ (left panel),
        and of $\lgeps{Be}$ for the disc-centre ($\mu=1.0$) 
        and limb ($\mu=0.2$) intensities (right panel).
        The 1D calculations adopt a fixed microturbulence
        of $1\,\kms$.}
        \label{fig:abcor}
    \end{center}
\end{figure*}

\begin{figure*}
    \begin{center}
        \includegraphics[scale=0.65]{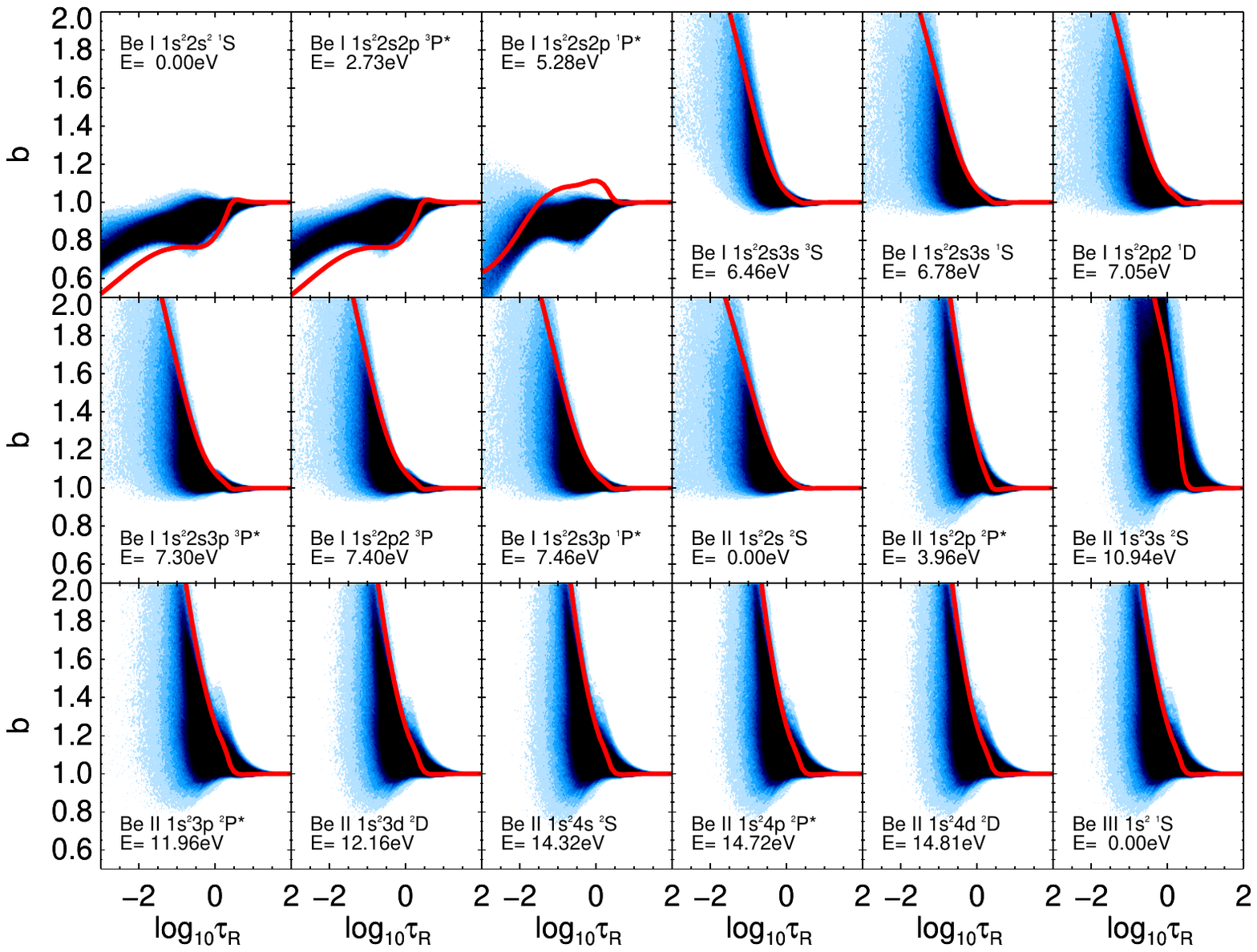}
        \caption{Departure coefficients for the lowest
        levels of neutral and ionised beryllium,
        and for the ground state of doubly ionised beryllium,
        as a function of the
        logarithmic Rosseland mean optical depth.
        Contours show the distribution of results from the 3D
        calculations; the thick red line
        shows the results from the 1D calculations.
        These calculations adopt $\lgeps{Be}=1.18$.}
    \label{fig:departure}
    \end{center}
\end{figure*}

\begin{figure*}
    \begin{center}
        \includegraphics[scale=0.325]{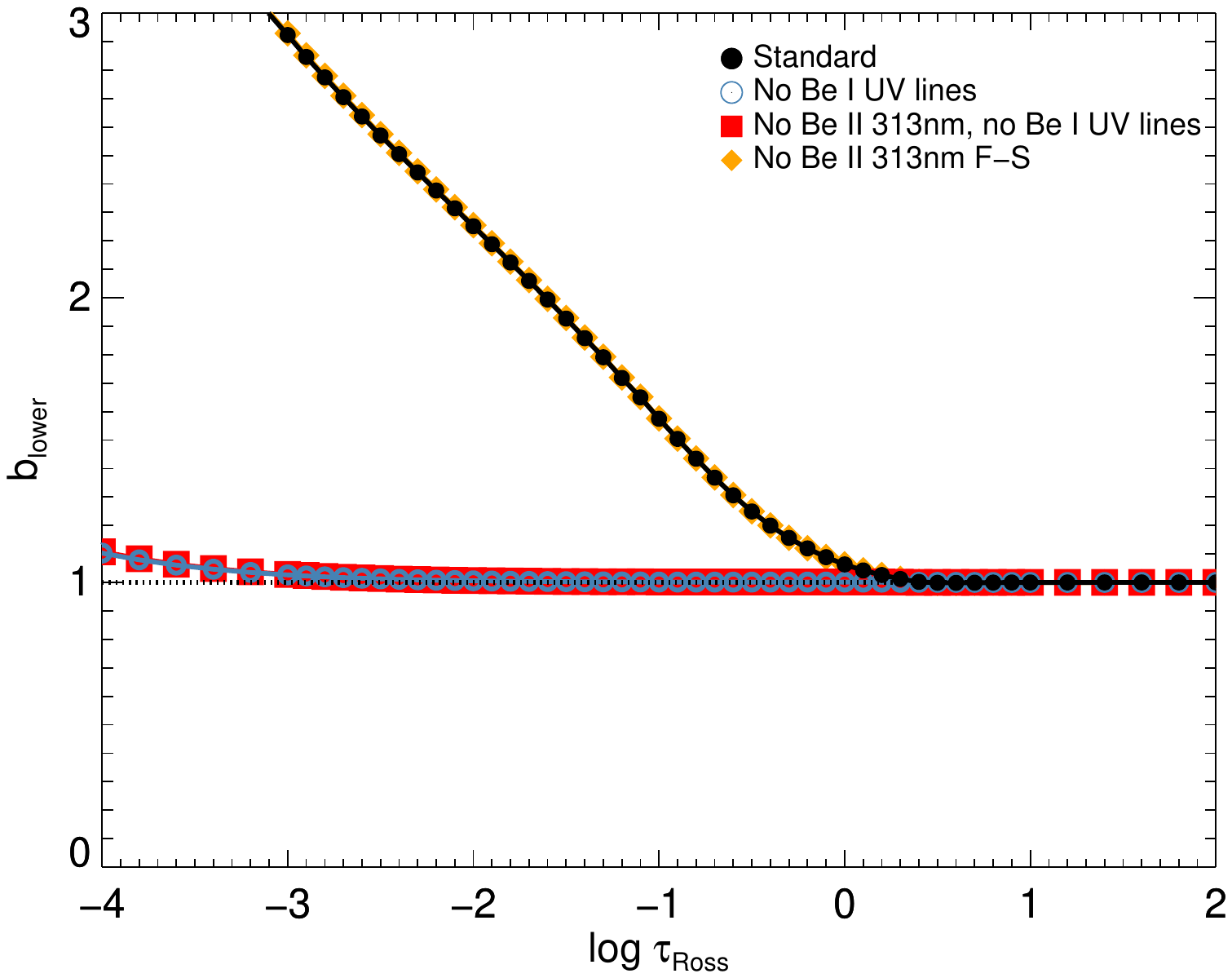}
        \includegraphics[scale=0.325]{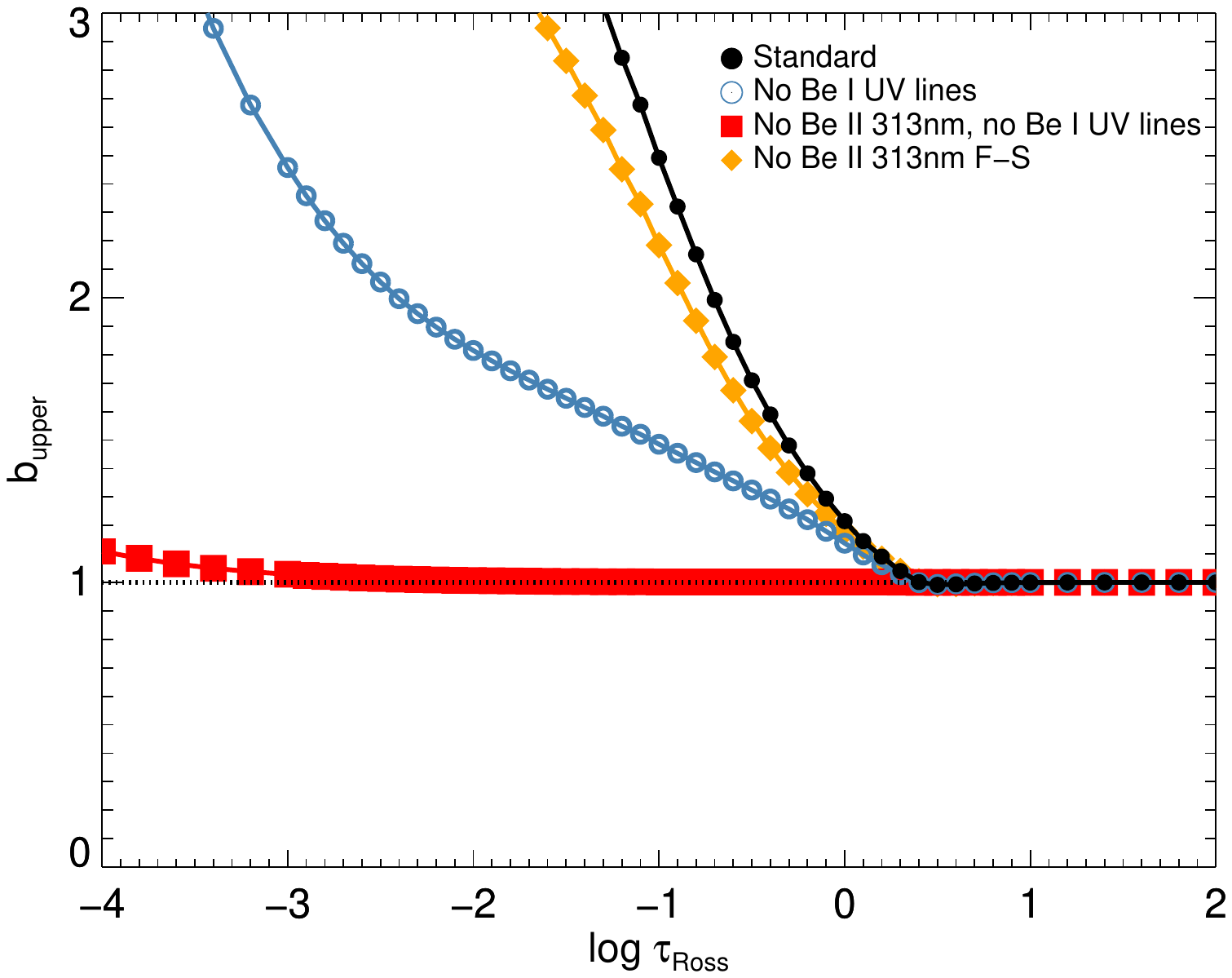}
        \includegraphics[scale=0.325]{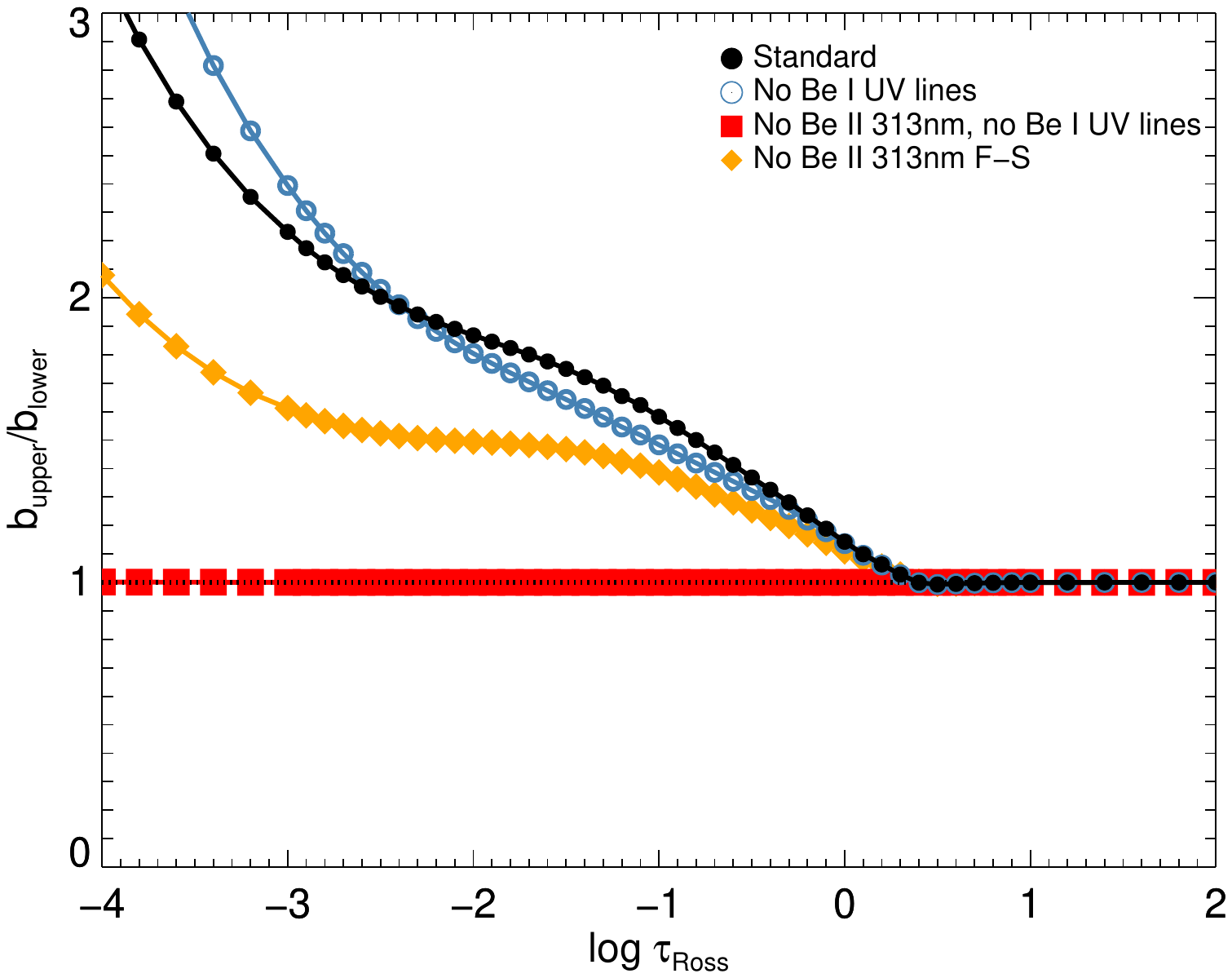}
        \includegraphics[scale=0.325]{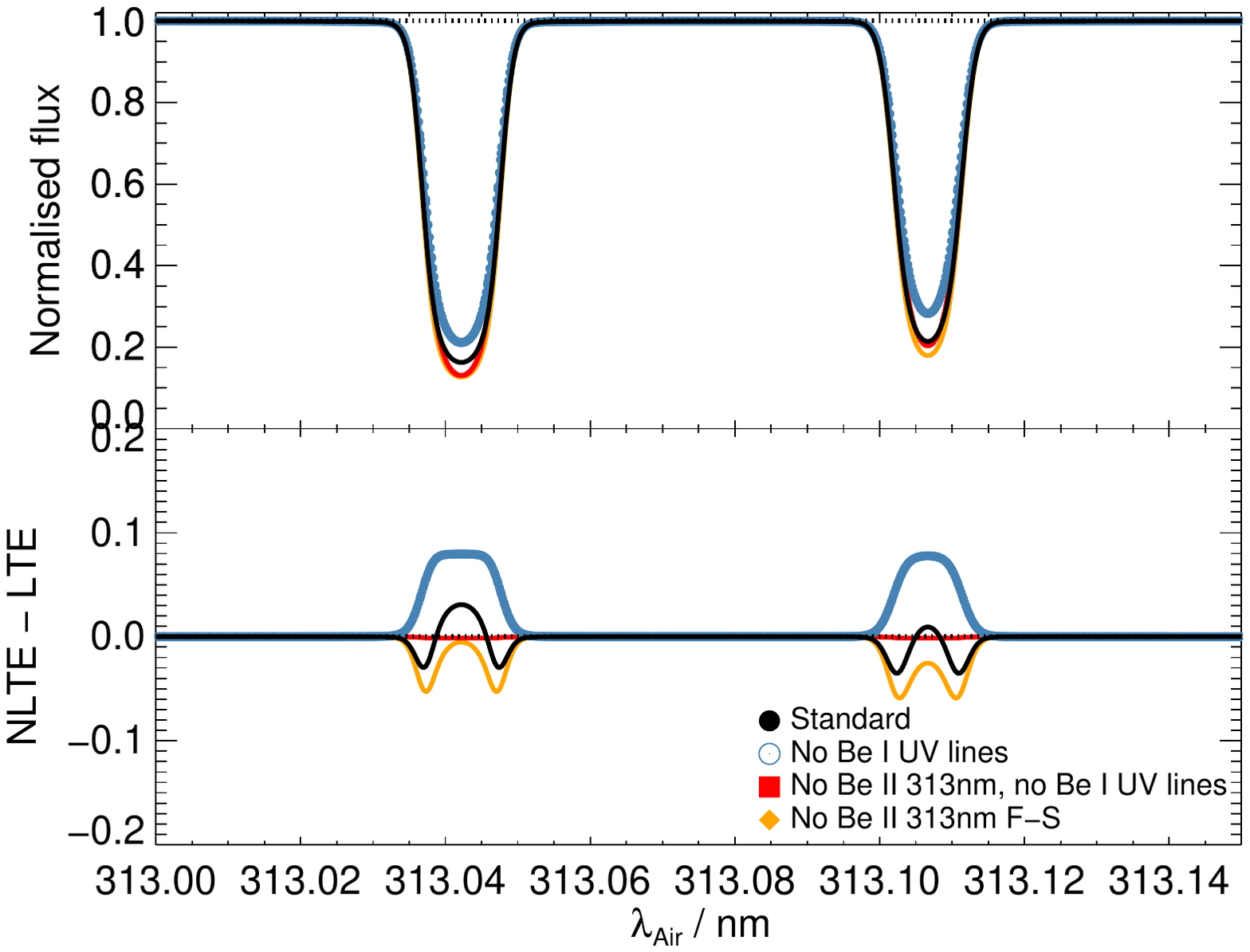}
        \caption{1D non-LTE effects on populations and emergent
        fluxes. Departure coefficients
        for the lower (upper left panel) and upper (upper right panel) levels
        of the \ion{Be}{II} $313\,\nm$ lines as a function
        of the logarithmic Rosseland mean optical depth, for
        the standard model (black circles and lines),
        as well as with different
        ingredients in the model atom switched off in
        the statistical equilibrium calculations:
        the standard model without any \ion{Be}{I} UV lines
        (blue open circles and lines); the same, but also
        without the \ion{Be}{II} $313\,\nm$ lines
        (red squares and lines); and the standard
        model without fine structure splitting of the 
        \ion{Be}{II} $313\,\nm$ lines
        considered in the calculation 
        of $J_{\mathrm{\nu}}$ and subsequently when evaluating
        the radiative rates 
        (No \ion{Be}{II} $313\,\nm$ F-S; orange diamonds and lines).
        Also shown are ratios of upper to lower
        departure coefficients (lower left panel), and
        the effects on the 
        \ion{Be}{II} $313\,\nm$ lines as viewed
        in the emergent disc-integrated flux albeit 
        without rotational broadening or macroturbulence (lower right panel).
        These calculations are based on the 
        1D \marcs{} model atmosphere and $\lgeps{Be}=1.18$.}
        \label{fig:nlte}
    \end{center}
\end{figure*}

\subsection{Model solar atmospheres}
\label{methodatmosphere}

The post-processing
radiative transfer calculations presented in this work were
performed on a 3D radiation-hydrodynamics simulation
of the solar surface carried out with the 
\stagger{} code \citep{2011JPhCS.328a2003C,2013A&A...557A..26M}.
This model atmosphere was first used 
and discussed in \citet{2018A&A...616A..89A},
and has been employed in several
follow-up studies \citep{2019A&A...624A.111A,2020A&A...636A.120A,
2021A&A...653A.141A};
in particular, it was used in the analysis
of molecular lines of carbon, nitrogen, and oxygen
presented in \citet{2021A&A...656A.113A}.
It was constructed assuming the solar composition presented in
\citet{2009ARA&A..47..481A}.
The mean effective temperature of the model
is $5773\,\mathrm{K}$, with a snapshot-to-snapshot
standard deviation of $16\,\mathrm{K}$.

Radiative transfer calculations were 
also performed on the theoretical \marcs{} 1D model 
of the solar atmosphere \citep{2008A&A...486..951G}.
These calculations were only used to quantify the 3D 
versus 1D effects, as well as to help elucidate the 
physical mechanisms of the departures from LTE,
as we discuss in \sect{methodeffects}.

\subsection{Spectrum synthesis}
\label{methodspectrum}

Two codes were used for the post-processing
radiative transfer calculations:
the 3D LTE code \scate{} \citep{2011A&A...529A.158H},
and the 3D non-LTE code \balder{} \citep{2018A&A...615A.139A}.
The latter code is based on \multitd{}
\citep{2009ASPC..415...87L_short}, with updates including
to the equation of state and opacities
\citep[e.g.][]{2016MNRAS.463.1518A,2023A&A...677A..98Z}.

The \scate{} calculations were performed on ten snapshots of the solar
model: this is the number of snapshots
recommended by \citet{2024A&A...688A.212R}
for reliable time-averaged spectral line profiles.
These calculations include the \ion{Be}{II} line, 
as well as $74$ known blending lines 
plus one hypothetical line, as we discuss in \sect{resultsblend}.
The abundances of background species were for the most part
set to those given in \citet{2021A&A...653A.141A}.
The exceptions are for carbon, nitrogen,
and oxygen: as molecules of CN and OH are present
in the \ion{Be}{II} $313.107\,\nm$ region,
the abundances of these elements were
set to those derived from a 3D LTE analysis of molecular lines given in
\citet{2021A&A...656A.113A}.
The spectra were calculated at disc-centre $\mu=1.0$
and at the limb $\mu=0.2$: the latter, via 
integration in the azimuthal angle $\phi$
using the $8$-point trapezoidal quadrature.
The disc-integrated fluxes were calculated using
the $4$-point Gauss quadrature for the $\mu$ integration
and the $4$-point trapezoidal quadrature for the $\phi$ integration
on the unit hemisphere.

As we discuss in \sect{results}, the analysis
had several free parameters including
the beryllium abundance, the oscillator strength and the wavelength of 
the hypothetical line that blends the \ion{Be}{II} $313.107\,\nm$ feature,
and a scale factor for the missing continuous opacity.
As such \scate{} calculations were performed to generate
a rectilinear grid of spectra, with each of 
these parameters taking a range of values.
The beryllium abundance and the oscillator strength of the blend
were varied in steps of $0.2\,\dex$,
the wavelength of the blend was varied in steps of $0.0005\,\nm$,
and the missing continuous opacity factor was varied in steps
of $0.25$.

The 3D non-LTE spectra were constructed by calculating 3D non-LTE to 3D LTE
ratios for the \ion{Be}{II} line using \balder{} on eight snapshots of the 
same solar model, 
and applying them to the blended 3D LTE spectrum calculated using
\scate{} as described above.  These 3D non-LTE calculations were carried out in
an analogous way to previous works on the Sun
\citep[e.g.][]{2018A&A...616A..89A,2019A&A...624A.111A,2020A&A...636A.120A}.
In particular, the mean radiation field $J_{\nu}$
was calculated using
$26$ rays on the unit sphere ($13$ outgoing, and $13$ ingoing),
including the two vertical rays, using the $8$-point Lobatto
quadrature for the $\mu$ integration and the $4$-point trapezoidal
quadrature for the $\phi$ integration.
The statistical equilibrium calculations were performed using the reduced model
atom that we discuss in \sect{methodatom}.  Although background line opacities
were included in the calculation of the statistical equilibrium, the final line
profiles that were used for the 3D non-LTE to 3D LTE ratios
were calculated without any blends. 
The emergent intensities were calculated
using the $10$-point Lobato quadrature
for the $\mu$ integration and 
the $8$-point trapezoidal quadrature for the $\phi$
integration on the unit hemisphere.
The vertical was explicitly included in this set,
but the rays at $\mu=0.2$ are not.
Consequently the 3D non-LTE to 3D LTE
ratios at the limb were found by cubic spline interpolation in $\mu$.
As with the \scate{} calculations, spectra were
calculated for a range of beryllium abundances in steps
of $0.2\,\dex$. In terms of equivalent width,
the 3D LTE profiles for the \ion{Be}{II} $313.107\,\nm$ 
line alone calculated by \scate{}
and \balder{} with the adopted setup
were found to agree to better than $0.005\,\dex$
for the disc-centre intensity and disc-integrated
flux, and $0.018\,\dex$ for the intensity at $\mu=0.2$.
The larger difference at the limb is mainly
due to the different treatment of UV continuous
opacity between the two codes,
which to first order cancels out in the 3D non-LTE 
to 3D LTE ratios used here.

Solely for the purpose of discussing
3D and non-LTE effects (\sect{methodeffects}),
calculations were also performed
on 1D model atmospheres (see \sect{methodatmosphere}).
\scate{} and \balder{} were also used for these
calculations, in a similar way as for the 3D calculations.
The main difference here is that a depth-independent microturbulence
of $1.0\,\kms$ was assumed.
No such microturbulence was adopted for the 3D calculations,
since these broadening effects are naturally accounted
for in the 3D radiative transfer calculations.
The discussion in \sect{methodeffects}
is based on equivalent widths, and
therefore macroturbulence (another free parameter in 
commonly used in analyses based on 1D models) is not 
relevant here.

\subsection{Model atom}
\label{methodatom}

A comprehensive model atom was constructed for this work,
for the most part following the steps described in 
Section 2.2 of \citet{2022A&A...657L..11K}.
The most significant difference 
is that, here, fine structure is taken into account
for the \ion{Be}{II} $313\,\nm$ lines
in the calculation of $J_{\mathrm{\nu}}$ and subsequently when evaluating
the radiative rates. In addition, recent data for 
inelastic collisions between ionised beryllium
and electrons from \citet{2024ADNDT.15601634D}
are included in this work.
The complexity of this comprehensive model was then reduced,
to make it feasible to run the 3D non-LTE calculations.
Early versions of 
these atoms are described in more detail in \citet{Ogneva1766090}.

We illustrate the grotrian diagram of the comprehensive model
in the left panels of \fig{fig:atom}.
This comprehensive model contains $236$ energy levels:
$181$ are of neutral beryllium,
$54$ are of ionised beryllium,
and the ground state of doubly ionised beryllium is also included.
The data come from the Atomic Spectra Database
of the National Institute of Standards and Technology
(NIST ADS; \citealt{2020Atoms...8...56R}) with
fine structure collapsed;
and from the Opacity Project Atomic Databse
(TOPBase; \citealt{1993A&A...275L...5C}).
These included levels 
reaching $0.14\,\mathrm{eV}$ below the ionisation
limit for neutral beryllium, and
$0.55\,\mathrm{eV}$ below the ionisation limit 
for ionised beryllium.
Rydberg levels missing in these data set
were then introduced, assuming them to be hydrogen-like;
however, unlike the other levels,
these were only allowed to couple collisionally
to the other levels in the model atom.

The photoionisation cross-sections as well as most of the 
bound-bound transition data originated from TOPBase.
The natural broadening parameters were calculated from the TOPBase data,
and broadening due to elastic collisions with neutral hydrogen
were calculated with the 
commonly used Lindholm-Foley-Uns\"{o}ld
theory (LFU, \citealt{1955psmb.book.....U};
see also \citealt{2016A&ARv..24....9B}),
with an enhancement factor of two.
The UV lines of \ion{Be}{I} as well
as the \ion{Be}{II} $313\,\nm$ lines themselves
strongly impact the statistical equilibrium
and were given special treatment:
the TOPBase oscillator strengths were replaced with those
from NIST ASD (originating from
\citealt{1999JPhB...32.5805T} and related calculations).
Moreover, the hydrogen broadening for these lines were
calculated using ABO theory
\citep{1995MNRAS.276..859A,1997MNRAS.290..102B,1998MNRAS.296.1057B,
2000MNRAS.311..535B}, taking the parameters listed in VALD.
In addition, fine structure was taken into account for these lines
in the calculation of $J_{\mathrm{\nu}}$ and subsequently when evaluating
the radiative rates and solving for the statistical equilibrium.
At least in this case (for the Sun), the effect of splitting the 
\ion{Be}{II} $313\,\nm$ lines in this way has 
a subtle impact on the equivalent widths, with a corresponding 
significant effect
on the abundance corrections because the lines are saturated,
as we discuss in \sect{effectsnlte}; 
whereas accounting for fine structure of the \ion{Be}{I} lines 
did not make a significant impact on the overall results.

The cross-sections for inelastic 
collisions with electrons (excitation and ionisation) 
involving the important low-lying levels of neutral and ionised beryllium
were taken
from \citet{2019ADNDT.127....1D} and \citet{2024ADNDT.15601634D}.
These data are primarily based on the convergent close-coupling
method (CCC; \citealt{1992cccc.rept.....B}); for neutral beryllium
data based on the B-Spline R-matrix method 
(BSR; \citealt{2006CoPhC.174..273Z}).
These data were supplemented with data from the 
ADAS project \citep{2011AIPC.1344..179S}.
For highly excited levels the inelastic electron collisions were
described with the recipes of 
\citet{1962ApJ...136..906V} and \citet{1973asqu.book.....A};
these are of lower accuracy, but through 1D non-LTE calculations
we verified that switching
them off had no impact on the strength of 
the \ion{Be}{II} $313.107\,\nm$ line.

For inelastic collisions with neutral hydrogen
(excitation and charge transfer),
the low-lying levels of neutral beryllium were described
using the data given in \citet{2016A&A...593A..27Y}.
These are based on an asymptotic model 
for the potentials combined with the Landau-Zener 
approach for the collision dynamics
\citep[e.g.][]{2013PhRvA..88e2704B,2016PhRvA..93d2705B}.
Data calculated using the free electron approach of 
\citet{1985JPhB...18L.167K,kaulakys1986free,1991JPhB...24L.127K},
in the scattering-length regime, were added to this
as motivated in
\citet{2018A&A...616A..89A,2019A&A...624A.111A}, 
and also discussed further in \citet{2024PhRvA.109e2820S}.
For ionised beryllium, excitation and ionisation
by impact with neutral hydrogen
were described using the Drawin recipe 
(as described in Appendix A
of \citealt{1993PhST...47..186L}), without any scaling.
Similarly to the tests on the electron rates,
we verified that switching off the free electron
rates and the Drawin rates did not impact
the \ion{Be}{II} $313.107\,\nm$ line in 1D non-LTE.

A reduced model atom was constructed from this
comprehensive one
in order to make the 3D non-LTE radiative transfer calculations feasible
(e.g.~Section 2.2.4 of \citealt{2024ARA&A..62.....L}).
Levels above $\mathrm{1s2.2s.5s\,^{1}S}$ of neutral
beryllium and $\mathrm{1s2.4f\,^{2}F^{o}}$ of ionised
beryllium were collapsed into ten and four super levels,
respectively.  The radiative and collisional transitions
involving these were similarly collapsed into super transitions.
We illustrate the grotrian diagram of this reduced model
in the right panels of \fig{fig:atom}.
Once again, we verified in 1D non-LTE that this reduction
had a negligible impact on the 
\ion{Be}{II} $313.107\,\nm$: at most $0.00014\,\dex$
in terms of abundance corrections, with the effects
largest at the solar limb.

\subsection{3D non-LTE effects}
\label{methodeffects}

We illustrate 
the 1D non-LTE, 3D LTE, and 3D non-LTE effects on abundances inferred from the
\ion{Be}{II} $313.107\,\nm$ line in \fig{fig:abcor}.
The panels illustrate various types of 
abundance corrections based on equivalent widths,
calculated at a given reference abundance.
For example, a positive value of the 3D non-LTE versus 1D LTE
abundance correction indicates that the reference
abundance inferred from a 1D LTE analysis of equivalent widths
underestimates the beryllium abundance compared to 
a 3D non-LTE analysis; the 1D LTE lines
in that case are too strong at given abundance.
The left panel shows the abundance corrections as a function
of $\mu$ for a fixed reference beryllium abundance
$\lgeps{Be}=1.18$, while the right panel shows 
the abundance corrections as a function of $\lgeps{Be}$,
at disc-centre and at the limb.

All of the abundance corrections in the intensities shown
in the left panel of \fig{fig:abcor} change sign when 
going from disc-centre to the limb.
A consequence of this is that the corresponding abundance corrections 
for the disc-integrated flux are close to zero.
The non-LTE effect on the disc-integrated flux (3D non-LTE versus
3D LTE abundance correction) amounts to $-0.027\,\dex$,
while the 3D effect (3D non-LTE versus 1D non-LTE abundance correction)
amounts to $+0.015\,\dex$.
The 1D non-LTE versus 1D LTE abundance correction is $-0.014\,\dex$
and consequently 
the 3D non-LTE versus 1D LTE abundance correction is close
to zero, at this abundance.
The 3D LTE versus 1D LTE abundance correction is somewhat larger
than the 3D non-LTE versus 1D non-LTE one:
$+0.031\,\dex$.

\subsubsection{Non-LTE effects}
\label{effectsnlte}

In order to gain a qualitative understanding
of the main non-LTE mechanism, it is helpful 
to look at the departure coefficients.
\fig{fig:departure} shows that,
for most levels (including the lower and upper level
of the \ion{Be}{II} $313$ lines), the departure coefficients
predicted by the 1D non-LTE calculations
roughly follow the distribution of departure coefficients
predicted by the 3D non-LTE calculations.
As such,
we have used 1D non-LTE calculations to 
study and understand the departures from LTE.
Quantitatively, 
the 3D non-LTE versus 3D LTE abundance corrections
are indeed similar to the 1D non-LTE versus 1D LTE abundance
corrections at disc-centre (\fig{fig:abcor});
the two quantities drift apart from each other towards the limb,
with more severe positive non-LTE corrections in the 3D model.

In the current model, the
main non-LTE effects relevant to the \ion{Be}{II} $313\,\nm$
lines
are caused by photon pumping of bound-bound transitions in the UV.
We illustrate this in \fig{fig:nlte}, using the results
of 1D non-LTE test calculations.
There are two competing photon pumping effects.
First, pumping of the
\ion{Be}{I} lines in the UV leads to an overpopulation of the excited levels of
neutral beryllium. This excess population flows into the ground state of ionised
beryllium via collisions.  
This increase in population leads to strengthening of
the \ion{Be}{II} $313\,\nm$ lines, an effect that
corresponds to negative abundance corrections relative to LTE.
Secondly, and in competition,
pumping of the \ion{Be}{II} $313\,\nm$ lines themselves tends to increase the
population of the upper level and reduce the population of the lower level.
This weakens the lines, corresponding to positive 
abundance corrections relative to LTE.
In \fig{fig:nlte} it can be seen that switching the 
\ion{Be}{I} UV lines off sets the departure coefficients of the lower
level of the \ion{Be}{II} $313\,\nm$ lines
(the ground state
of ionised beryllium) to be close to unity (upper left panel
of \fig{fig:nlte});
and further switching off the \ion{Be}{II} $313\,\nm$ lines themselves
then also sets the departure coefficients of the 
upper level to be close to unity (upper right panel of \fig{fig:nlte});
the emergent line profiles in that case then do not display
any deviations from LTE
(lower right panel of \fig{fig:nlte}).

This picture is qualitatively similar to that
presented in Section 3.4 of \citet{1995A&A...297..787G}. The main difference is
that those authors find a significant pumping of the \ion{Be}{I} photoionisation
processes themselves; here, it is rather the pumping of the \ion{Be}{I} lines
that are driving the effects.
Quantitatively, the 
1D non-LTE versus 1D LTE abundance corrections for the 
disc-integrated flux ($-0.014\,\dex$ as we discussed above) 
are similar to those
given by \citet{1995A&A...297..787G} and \citet{2011PASJ...63..697T}.
\citet{2022A&A...657L..11K} report a more severe correction of $-0.07\,\dex$.
As we illustrate in \fig{fig:nlte}, this is possibly
due to their neglecting fine structure of the
\ion{Be}{II} $313\,\nm$ lines in their
statistical equilibrium calculations
(when calculating $J_{\nu}$ and
subsequently evaluating the radiative rates).
Merging the lines reduces
the photoexcitation rates in the transition (see the discussion in Appendix B of
\citealt{2015A&A...583A..57S}).  Thus by doing so the second non-LTE effect
described above (pumping of the \ion{Be}{II} $313\,\nm$ lines themselves, which
weakens the lines and leads to positive 1D non-LTE abundance corrections)
is reduced and the lines get stronger overall (``No \ion{Be}{II} $313\,\nm$ 
F-S'' in
the lower right panel of
\fig{fig:nlte}).  In this model, assuming $\lgeps{Be}=1.18$, the equivalent
width then increases such that the 1D non-LTE versus 1D LTE 
abundance correction becomes $-0.07\,\dex$, in excellent
agreement with \citet{2022A&A...657L..11K}.

The competition between the two non-LTE effects helps explain the 
behaviour of the 3D non-LTE versus 3D LTE 
abundance corrections with disc position
and with 3D LTE beryllium abundance
seen in \fig{fig:abcor} (red diamonds).  These abundance
corrections are negative at disc-centre, suggesting that photon pumping of the
\ion{Be}{I} UV lines is more important there.  At the limb, the 3D non-LTE
versus 3D LTE 
abundance corrections are positive, suggesting that instead photon
pumping of the \ion{Be}{II} $313\,\nm$ lines themselves increasingly 
dominate at higher layers in the atmosphere.
This is consistent with the run of the ratio
of departure coefficients of the upper and lower
levels (lower left panel of \fig{fig:nlte}):
this ratio increases towards higher layers, indicating an increasing
line source function relative to LTE and therefore more
weakening relative to LTE.
At the limb, the 3D non-LTE versus 3D LTE 
abundance corrections show a strong sensitivity
to the beryllium abundance, becoming more negative
with increasing $\lgeps{Be}$.
Increasing the beryllium abundance increases the strength of the
\ion{Be}{II} $313\,\nm$ lines, up to and beyond saturation; 
this in turn increases
the photon losses in the line (which is efficient for saturated lines), 
which counteracts the photon-pumping effect described above.

\subsubsection{3D effects and their coupling with the non-LTE effects}
\label{effects3d}

The 3D effects on the \ion{Be}{II} $313\,\nm$ lines
are here quantified 
by the 3D non-LTE versus 1D non-LTE abundance corrections.
These are slightly positive at disc-centre (blue squares in \fig{fig:abcor}).
This is consistent with findings for other
saturated resonance lines of majority species,
as seen in Figure 6 of 
\citet{2024ARA&A..62.....L},
where the effects are separated
into that caused by differences in the mean stratification and
that caused by inhomogeneities resulting from solar granulation.
According to that plot, the effect of the mean stratification dominates: 
the 3D model has a slightly shallower temperature gradient, which 
weakens the lines compared to those generated from the 1D model.

At the limb, the 3D non-LTE versus 1D non-LTE abundance corrections
change sign, becoming negative and more severe: 
the 3D non-LTE line is much stronger than the 
1D non-LTE one, at given 1D non-LTE abundance.
When observing the limb, the granules are seen edge-on.
The horizontal velocity fields being larger than 
the vertical velocity fields in the 3D models
(e.g.~\citealt{1998ApJ...499..914S}, Figure 5),
the spectral lines experience more broadening.
This de-saturates the 
\ion{Be}{II} $313\,\nm$ lines, strengthening them
in the 3D calculations.
In the 1D calculations, this effect can roughly be accounted for by
using a larger microturbulence towards the limb
(see the discussions in \citealt{2013MSAIS..24...37S} and
\citealt{2022SoPh..297....4T});
here, the 1D calculations assumed a fixed 
microturbulence of $1.0\,\kms$.

It is interesting to consider how
the 3D non-LTE versus 1D LTE abundance corrections
compare against the combination of the 
3D LTE versus 1D LTE abundance corrections
and the 1D non-LTE versus 1D LTE abundance corrections.
These are shown in \fig{fig:abcor}.
Although the 3D non-LTE versus 1D LTE abundance
corrections fall in between the two, they do not match
up exactly; this discrepancy reflects the coupling of the 3D effects
with the non-LTE effects. At the limb, at $\lgeps{Be}=1.18$, the 
3D non-LTE versus 1D LTE abundance correction is $-0.03\,\dex$.
The 3D LTE versus 1D LTE abundance correction is $-0.08\,\dex$
and the 1D non-LTE versus 1D LTE abundance correction is $0.09\,\dex$;
combining the two gives $+0.01\,\dex$,
which is $0.04\,\dex$ away from the 3D non-LTE result.
This discrepancy becomes larger at other abundances:
at very low abundances, $\lgeps{Be}=0.38$ for example,
the discrepancy reaches $0.09\,\dex$.

The coupling between the 3D effects and the non-LTE
effects can also be seen on the departure coefficients
upon closer inspection of \fig{fig:departure}.
In particular, the $\mathrm{1s^{2}{2s}{2p}\,^{1}P^{o}}$ level
of neutral beryllium (first row, third column) shows an overpopulation
around a logarithmic Rosseland mean optical depth of $-1$
in the 1D calculations; whereas this level
is systematically underpopulated in the 3D model
at those depths.

\section{Analysis of the solar beryllium abundance}
\label{results}

\begin{table*}
\begin{center}
\caption{Line list adopted for the synthesis of the \ion{Be}{II} $313.107\,\nm$ region. }
\label{tab:linelist}
\begin{tabular}{l c c c c | l c c c c}
\hline
\hline
\noalign{\smallskip}
Species & 
$\lambda_{\mathrm{Air}}/\nm$  & 
$E_{\mathrm{low}}/\mathrm{eV}$ & 
$\log{gf}$ & 
$\lgeps{X}$ &
Species & 
$\lambda_{\mathrm{Air}}/\nm$  & 
$E_{\mathrm{low}}/\mathrm{eV}$ & 
$\log{gf}$ & 
$\lgeps{X}$ \\
\noalign{\smallskip}
\hline
\hline
\noalign{\smallskip}
$^{\mathrm{a}}$\ion{Ti}{II} & $313.079869$ & $0.0117$ & $-1.190$ & $4.97$ & \ion{Mn}{I} & $313.103629$ & $3.7723$ & $-1.667$ & $5.42$ \\ 
\ion{Ce}{II} & $313.080457$ & $0.8085$ & $-1.830$ & $1.58$ & CN & $313.104785$ & $0.9476$ & $-6.251$ & $8.47/7.89$ \\ 
\ion{Ar}{I} & $313.080815$ & $11.5484$ & $-2.910$ & $6.38$ & \ion{Ar}{I} & $313.105415$ & $11.5484$ & $-3.780$ & $6.38$ \\ 
CN & $313.081074$ & $0.9612$ & $-2.987$ & $8.47/7.89$ & CN & $313.105501$ & $1.2900$ & $-6.588$ & $8.47/7.89$ \\ 
\ion{Gd}{II} & $313.081315$ & $1.1566$ & $-0.083$ & $1.08$ & \ion{Fe}{II} & $313.105884$ & $9.7002$ & $-3.717$ & $7.46$ \\ 
OH & $313.081359$ & $1.9392$ & $-3.782$ & $8.70/12.00$ & \ion{Mn}{II} & $313.105933$ & $6.6711$ & $-2.692$ & $5.42$ \\ 
\ion{Co}{II} & $313.081830$ & $9.2397$ & $-6.551$ & $4.94$ & \ion{V}{II} & $313.106325$ & $4.2438$ & $-4.386$ & $3.90$ \\ 
CN & $313.081830$ & $0.3326$ & $-4.245$ & $8.47/7.89$ & $^{\mathrm{c}}$\ion{Be}{II} & $313.106516$ & $0.0000$ & $-0.479$ & $1.21$ \\ 
\ion{Cu}{I} & $313.083154$ & $6.1920$ & $-6.851$ & $4.18$ & CN & $313.106737$ & $1.2900$ & $-6.195$ & $8.47/7.89$ \\ 
CN & $313.083154$ & $0.9612$ & $-6.261$ & $8.47/7.89$ & \ion{Th}{II} & $313.107016$ & $0.0000$ & $-1.559$ & $0.03$ \\ 
CN & $313.083889$ & $0.3326$ & $-4.272$ & $8.47/7.89$ & \ion{Fe}{II} & $313.109032$ & $11.2065$ & $-5.030$ & $7.46$ \\ 
\ion{V}{I} & $313.084164$ & $1.9553$ & $-3.762$ & $3.90$ & $^{\mathrm{d}}$\ion{Cr}{II} & $313.110022$ & $8.5981$ & $-2.636$ & $5.62$ \\ 
\ion{Cr}{II} & $313.084350$ & $10.7484$ & $-6.564$ & $5.62$ & $^{\mathrm{e}}$\ion{Fe}{II} & $313.110425$ & $9.6878$ & $-1.036$ & $7.46$ \\ 
\ion{Ce}{II} & $313.085183$ & $0.0000$ & $-3.010$ & $1.58$ & \ion{Fe}{I} & $313.111013$ & $3.0469$ & $-5.610$ & $7.46$ \\ 
\ion{Cu}{I} & $313.086997$ & $6.1227$ & $-2.728$ & $4.18$ & \ion{Zr}{I} & $313.111016$ & $0.5203$ & $-0.400$ & $2.59$ \\ 
\ion{Fe}{II} & $313.087056$ & $8.9593$ & $-5.404$ & $7.46$ & CO & $313.111229$ & $5.2067$ & $-6.164$ & $8.47/7.89$ \\ 
\ion{Ce}{II} & $313.087086$ & $1.0897$ & $-2.250$ & $1.58$ & \ion{Os}{I} & $313.111616$ & $1.8409$ & $+0.050$ & $1.35$ \\ 
\ion{Ce}{II} & $313.087586$ & $1.1069$ & $-0.320$ & $1.58$ & CN & $313.112102$ & $0.7697$ & $-3.352$ & $8.47/7.89$ \\ 
\ion{Cr}{II} & $313.088723$ & $9.5803$ & $-2.745$ & $5.62$ & \ion{V}{II} & $313.112307$ & $6.5807$ & $-5.242$ & $3.90$ \\ 
\ion{Cu}{I} & $313.090076$ & $6.1227$ & $-3.697$ & $4.18$ & \ion{Ti}{I} & $313.114298$ & $0.8360$ & $-6.733$ & $4.97$ \\ 
\ion{Fe}{II} & $313.090478$ & $7.4867$ & $-2.429$ & $7.46$ & CN & $313.114475$ & $0.7698$ & $-6.763$ & $8.47/7.89$ \\ 
CN & $313.092136$ & $0.2947$ & $-4.364$ & $8.47/7.89$ & CN & $313.114544$ & $1.0137$ & $-6.561$ & $8.47/7.89$ \\ 
\ion{Ce}{II} & $313.092185$ & $0.4954$ & $-0.760$ & $1.58$ & CN & $313.115515$ & $0.5037$ & $-5.142$ & $8.47/7.89$ \\ 
OH & $313.093341$ & $0.6832$ & $-3.359$ & $8.70/12.00$ & CN & $313.115613$ & $0.5037$ & $-4.841$ & $8.47/7.89$ \\ 
CN & $313.093518$ & $0.2947$ & $-4.397$ & $8.47/7.89$ & \ion{Mo}{I} & $313.119416$ & $2.4992$ & $-1.356$ & $1.88$ \\ 
CN & $313.095195$ & $0.0013$ & $-5.882$ & $8.47/7.89$ & CN & $313.119634$ & $0.3414$ & $-4.219$ & $8.47/7.89$ \\ 
CN & $313.095744$ & $0.0013$ & $-6.039$ & $8.47/7.89$ & \ion{Cr}{I} & $313.119918$ & $3.0106$ & $-6.270$ & $5.62$ \\ 
CN & $313.095950$ & $1.2986$ & $-6.176$ & $8.47/7.89$ & \ion{Ni}{II} & $313.120545$ & $12.7385$ & $-3.114$ & $6.20$ \\ 
\ion{Cr}{II} & $313.096950$ & $6.8027$ & $-4.643$ & $5.62$ & $^{\mathrm{f}}$\ion{Cr}{I} & $313.120692$ & $3.1128$ & $-0.401$ & $5.62$ \\ 
CN & $313.097264$ & $0.0127$ & $-5.531$ & $8.47/7.89$ & CN & $313.120947$ & $2.3116$ & $-5.598$ & $8.47/7.89$ \\ 
CN & $313.097411$ & $0.0127$ & $-5.594$ & $8.47/7.89$ & CN & $313.121781$ & $0.3414$ & $-4.245$ & $8.47/7.89$ \\ 
OH & $313.099275$ & $1.5694$ & $-4.030$ & $8.70/12.00$ & CN & $313.121860$ & $0.9345$ & $-3.005$ & $8.47/7.89$ \\ 
\ion{N}{II} & $313.099516$ & $20.4091$ & $-3.140$ & $7.83$ & CN & $313.122409$ & $2.3116$ & $-2.454$ & $8.47/7.89$ \\ 
CN & $313.099608$ & $0.9476$ & $-2.987$ & $8.47/7.89$ & CN & $313.123665$ & $1.2742$ & $-6.964$ & $8.47/7.89$ \\ 
\ion{Mn}{II} & $313.101520$ & $6.1113$ & $-1.230$ & $5.42$ & \ion{Fe}{I} & $313.124311$ & $2.1759$ & $-3.792$ & $7.46$ \\ 
$^{\mathrm{b}}$\ion{Ti}{II} & $313.102096$ & $0.0100$ & $-3.875$ & $4.97$ & CN & $313.125077$ & $0.9345$ & $-3.023$ & $8.47/7.89$ \\ 
CN & $313.102461$ & $0.9978$ & $-6.746$ & $8.47/7.89$ & CN & $313.125145$ & $2.3116$ & $-2.475$ & $8.47/7.89$ \\ 
CN & $313.102775$ & $0.9476$ & $-3.005$ & $8.47/7.89$ & \ion{Tm}{II} & $313.125515$ & $0.0000$ & $+0.240$ & $0.11$ \\ 
\noalign{\smallskip}
\hline
\hline
\end{tabular}
\end{center}
\tablefoot{Lines from the VALD database, with modifications noted below. Adopted
elemental abundances from \citet{2021A&A...653A.141A} for atomic and ionic
species, and \citet{2021A&A...656A.113A} for the different elements in molecular
species of C, N, and O. \tablefoottext{a}{ABO parameters calculated in this work
$\{\sigma,\alpha\}=\{217\,\mathrm{a_{0}^2},0.213\}$.}
\tablefoottext{b}{Hypothetical blend with fitted wavelength and oscillator
strength.} \tablefoottext{c}{Beryllium line with fitted abundance, and VALD ABO
parameters $\{\sigma,\alpha\}=\{123\,\mathrm{a_{0}^2},0.212\}$.}
\tablefoottext{d}{VALD ABO parameters
$\{\sigma,\alpha\}=\{519\,\mathrm{a_{0}^2},0.213\}$.} \tablefoottext{e}{VALD ABO
parameters $\{\sigma,\alpha\}=\{407\,\mathrm{a_{0}^2},0.284\}$.}
\tablefoottext{f}{Wavelength from \citet{1975A&A....42...37C}, and ABO
parameters calculated in this work
$\{\sigma,\alpha\}=\{415\,\mathrm{a_{0}^2},0.262\}$.}} 
\end{table*}

\begin{figure}
    \begin{center}
        \includegraphics[scale=0.325]{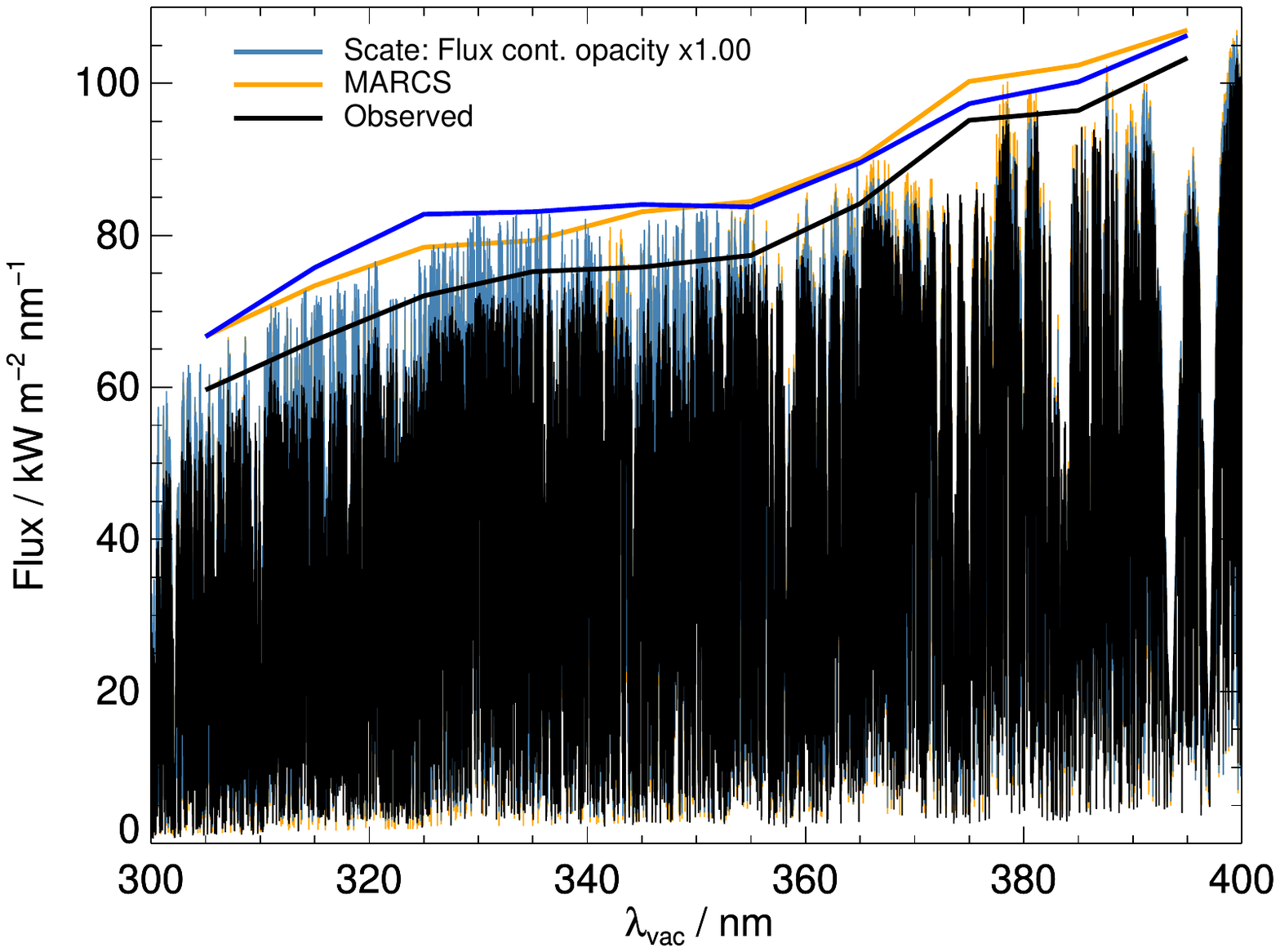}
        \includegraphics[scale=0.325]{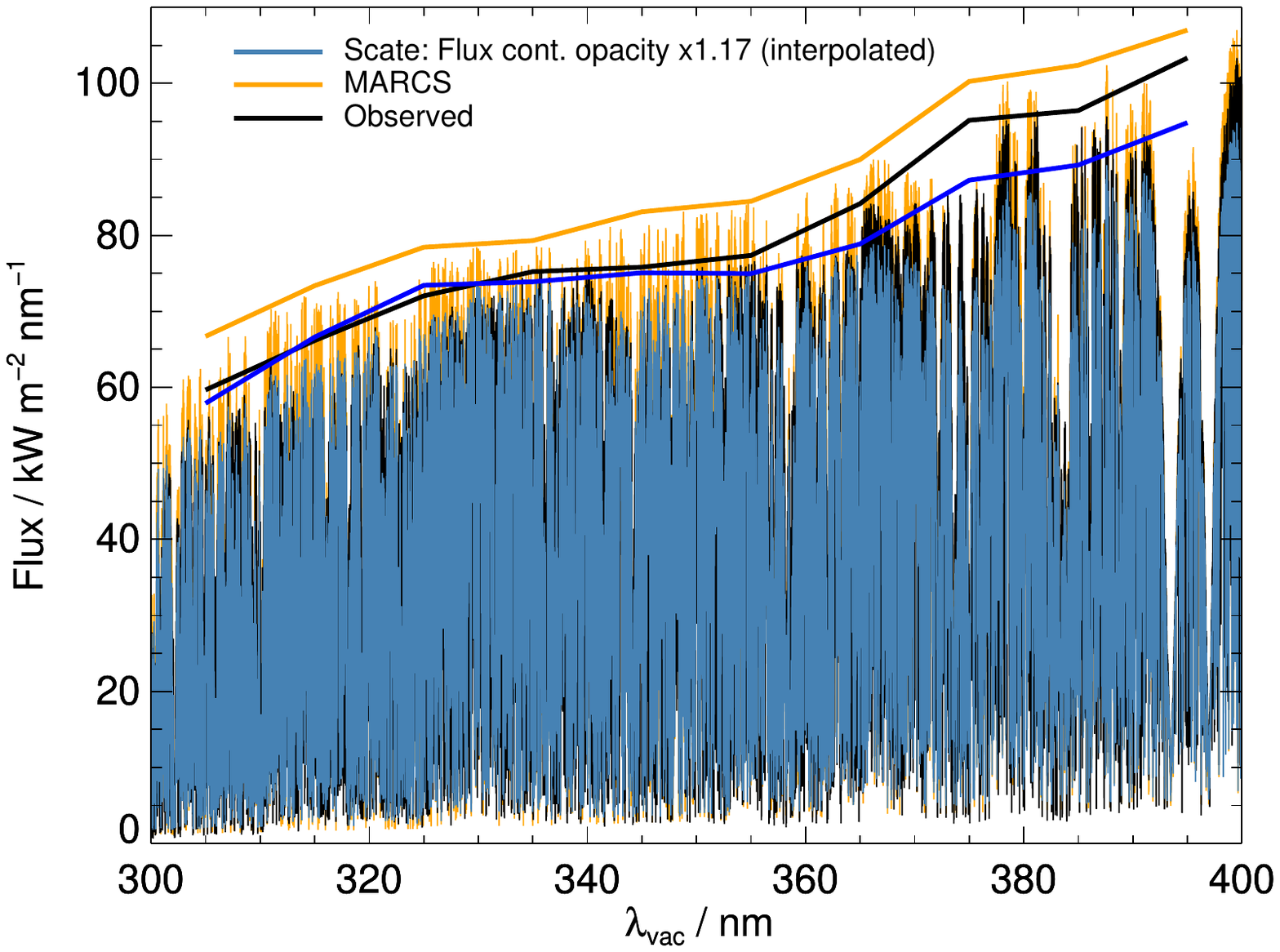}
        \includegraphics[scale=0.325]{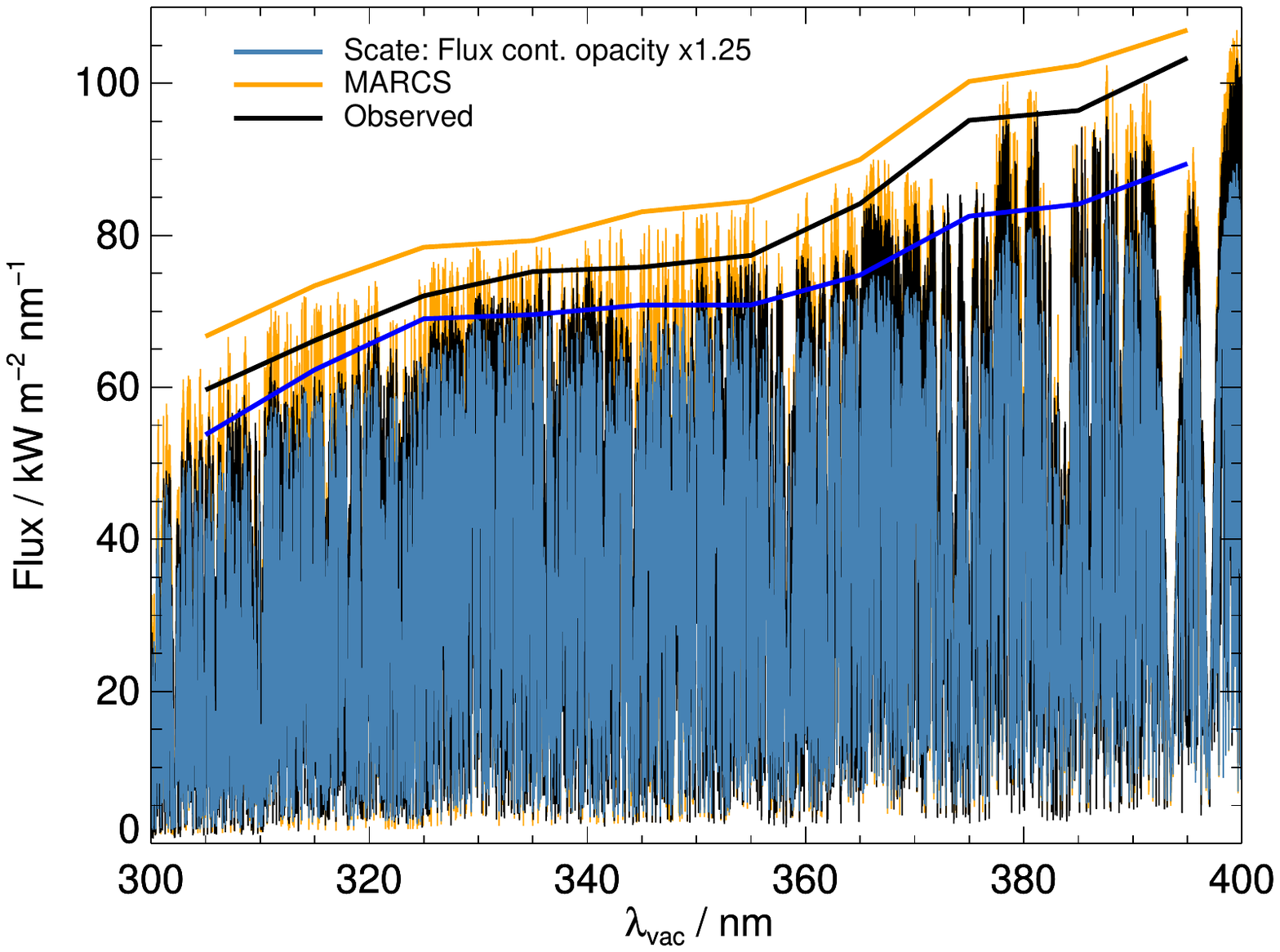}
        \caption{Disc-integrated flux at the solar surface.
        Two sets of theoretical data are shown:
        the \scate{} data are from this work,
        while the \marcs{} data are from \citet{2008A&A...486..951G}.
        Also shown are results derived
        from the solar irradiance atlas of 
        \citet{2005MSAIS...8..189K}.
        The \scate{} data are shown without any
        scaling of the continuous opacity (top panel),
        and with a factor of $1.25$ applied (bottom panel),
        as well as with a factor of $1.17$ applied 
        as determined by interpolation (middle panel).
        The continua are approximately traced by taking
        maxima within bins of widths of $10\,\nm$.}
        \label{fig:opacity}
    \end{center}
\end{figure}

\begin{figure}
    \begin{center}
        \includegraphics[scale=0.325]{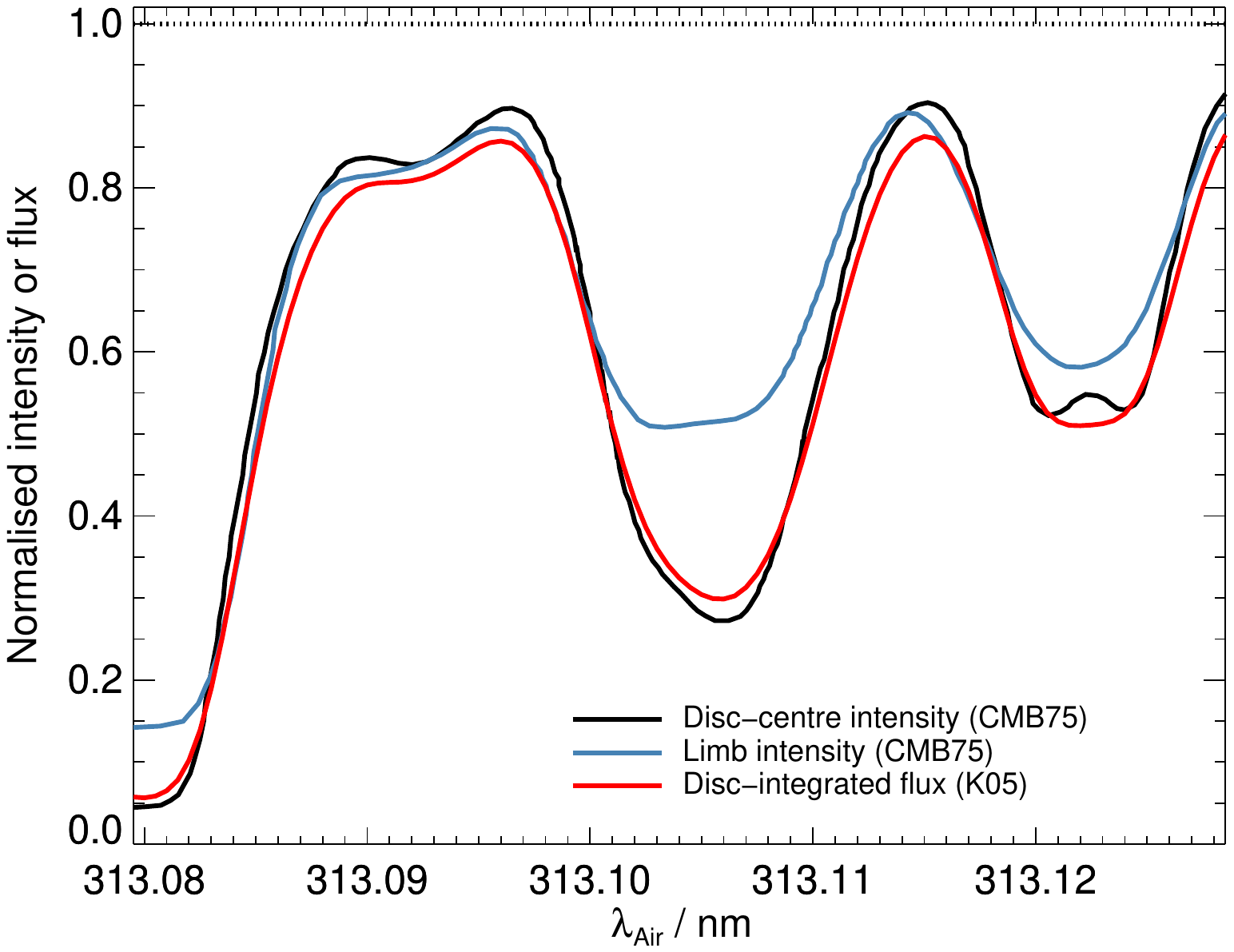}
        \caption{Continuum-normalised
        observational data used for spectral line fitting,
        over-plotted for comparison.
        Disc-centre and limb intensities from 
        \citet{1975A&A....42...37C}, and disc-integrated flux
        from \citet{2005MSAIS...8..189K}.
        All spectra are plotted as a function of wavelength in air
        and are already corrected for the solar gravitational redshift.}
        \label{fig:obs}
    \end{center}
\end{figure}

\begin{figure*}
    \begin{center}
        \includegraphics[scale=0.325]{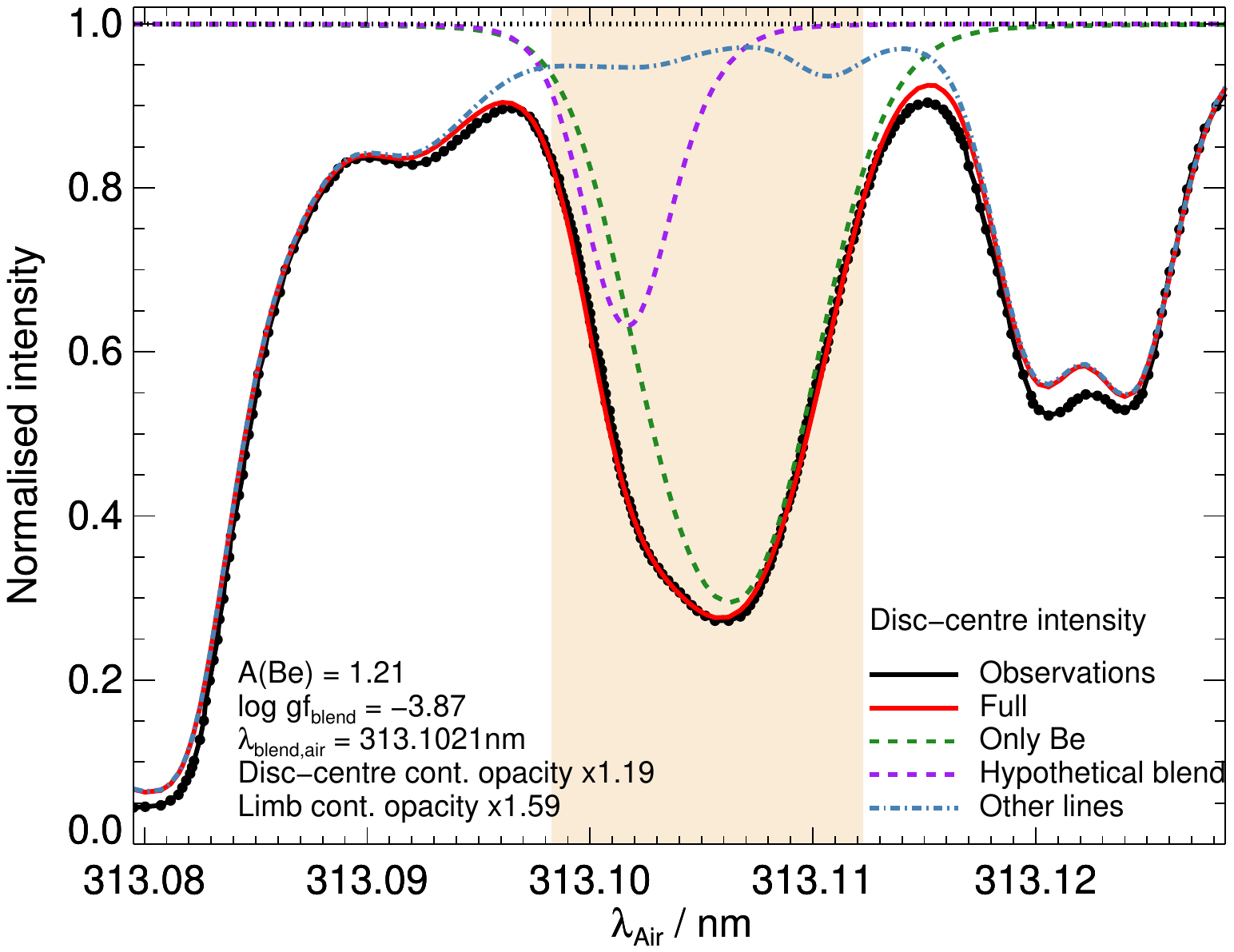}
        \includegraphics[scale=0.325]{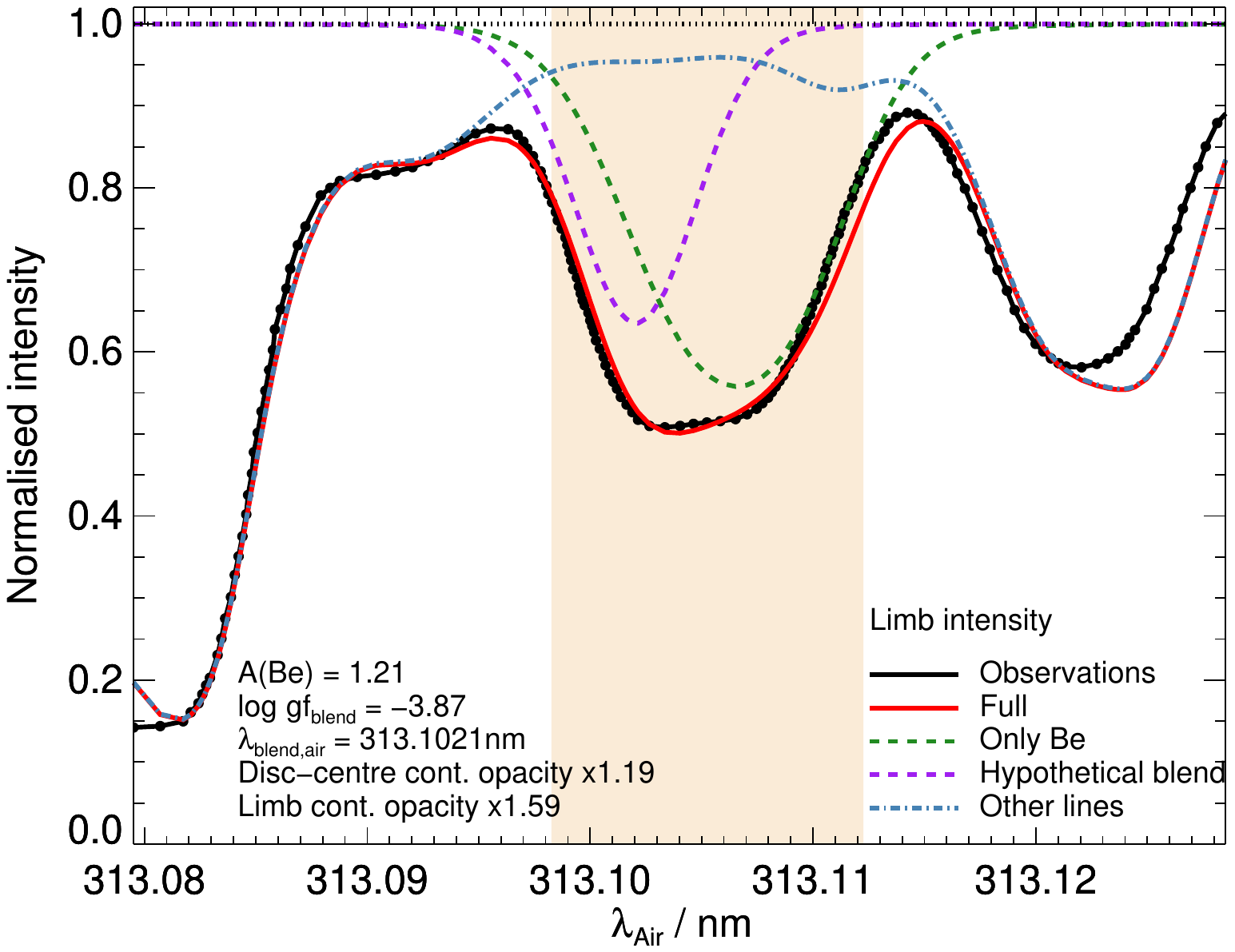}
        \caption{Fits to the disc-centre (left panel)
        and limb (right panel) intensities
        resulting from the calibration
        of the blend described in \sect{resultsblend}.
        The shaded area shows the wavelength range 
        considered in the fit.}
        \label{fig:blend}
    \end{center}
\end{figure*}

\begin{figure}
    \begin{center}
        \includegraphics[scale=0.325]{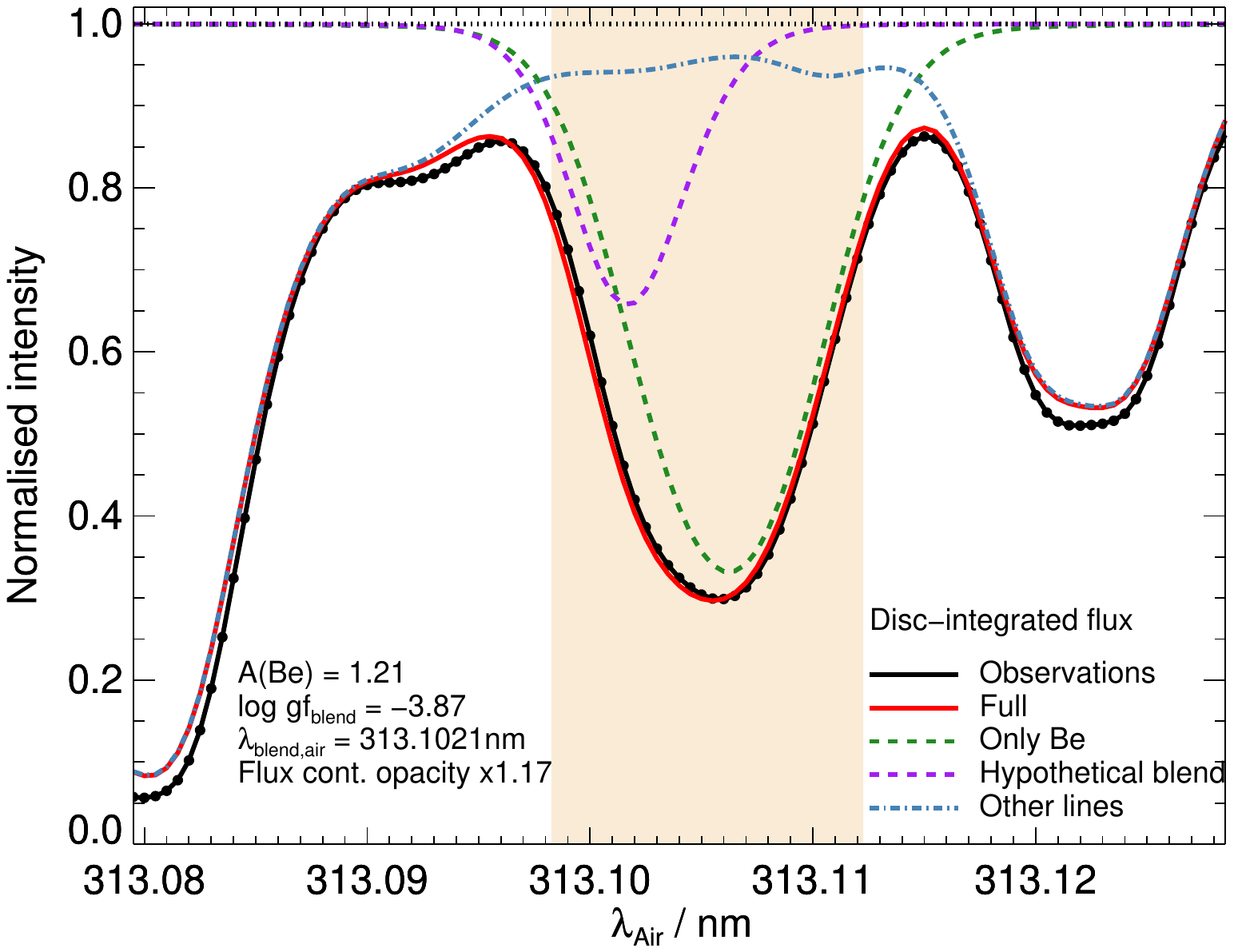}
        \caption{Fit to the disc-integrated flux
        described in \sect{resultsabundance},
        based on the 
        continuous opacity
        calibration in \sect{resultsopacity}
        and blend
        calibration in \sect{resultsblend}.
        The shaded area shows the wavelength range 
        considered in the fit.}
        \label{fig:flux}
    \end{center}
\end{figure}

The solar beryllium abundance was determined by fitting the
3D non-LTE models described in \sect{method}
to observed spectra. These fits proceeded in three steps.

In the first step (\sect{resultsopacity}), the effect
of missing continuous opacity 
on the theoretical disc-integrated flux was
calibrated. In the second step (\sect{resultsblend}),
the wavelength and oscillator strength of the blend at around $313.102\,\nm$
was calibrated on disc-centre and limb observations.
With these calibrations in hand,
in the third step (\sect{resultsabundance}) the 
\ion{Be}{II} $313.107\,\nm$ line was fit to the solar flux data to
obtain an
inferred value of the solar beryllium abundance and an uncertainty.
The spectral line fits were performed by $\chi^{2}$ minimisation
using \texttt{MPFIT} \citep{2009ASPC..411..251M}.

Each step is based on different
observational data presented in the literature.
We show the solar irradiance data (converted to 
solar flux at the stellar surface) in \fig{fig:opacity}, and the
continuum-normalised disc-centre and limb intensities
as well as disc-integrated flux in \fig{fig:obs}.
All of these data were
taken with the McMath-Pierce solar telescope 
at the Kitt Peak National Observatory,
via \citet{1975A&A....42...37C} and 
\citet{2005MSAIS...8..189K}.
The Fourier Transform Spectrograph has a nominal 
resolving power of around $R=3\times10^{5}$.

\subsection{Step 1: Missing continuous opacity calibration}
\label{resultsopacity}

The solar irradiance atlas presented by \citet{2005MSAIS...8..189K}
was used to calibrate the effect of the missing continuous opacity 
on the theoretical disc-integrated
flux.\footnote{\url{http://kurucz.harvard.edu/sun/irradiance2005/irradthu.dat}.}
We illustrate this in \fig{fig:opacity}.
It is clear that the continuum flux is slightly too high in the models
used in this work, as well as in 
for instance the fluxes released by
\citet{2008A&A...486..951G} for their solar model.
We also illustrate the results after applying
a scale factor of $1.25$ to the continuous opacity,
which is seen to be a slight overcorrection.
Interpolating between the two results,
the scale factor is roughly $1.17$ in 
the \ion{Be}{II} $313.107\,\nm$ region.
The uncertainty of this scaling factor was estimated to be $0.05$
based on different continuum placements,
and by itself corresponds to a $0.02\,\dex$ uncertainty
in $\lgeps{Be}$, and neglecting the missing
opacity entirely when fitting the 
disc-integrated flux would reduce the
inferred abundance by $0.065\,\dex$.

This method of calibration is inspired by that of \citet{2022A&A...657L..11K},
but here we make the point of using high-resolution data.
Low-resolution data are significantly
depressed by spectral lines, as shown
in the top panel of Figure 13 of \citet{2023A&A...677A..98Z} for example.
This would have made the calibration also dependent on a complete
and accurate description of spectral lines:
an incomplete line list would lead to an overestimation of
the missing continuous opacity.

An alternative, and independent, approach to calibrating the missing
continuous opacity is to enforce consistency
between oxygen abundances inferred from OH lines
in the UV and in the infrared
\citep{1998Natur.392..791B,2004A&A...417..769A}.
This approach is prone
to uncertainties in the oxygen abundance itself
\citep{2018ApJ...865....8C},
and to imprecision (and possibly
systematic biases) due to difficulties
in placing the continuum and 
in taking blends into account.
Moreover, any departures from LTE could affect
the UV lines more than the infrared lines 
(\citealt{1975MNRAS.170..447H}; see also
the discussion in Section 2.5.2 of 
\citealt{2024ARA&A..62.....L}), the effect of which
would conflate with that of the missing continuous opacity.
Nevertheless, it is reassuring that
the enhancement of $1.17$ derived here closely agrees with
the value of $1.2$ obtained in Section 3 of \citet{2018ApJ...865....8C} 
using the OH method, although this could be a coincidence
due to the different tools and atomic data being used.

The enhancement to the continuous opacity is expected
to have a centre-to-limb variation itself,
and the value of $1.17$ derived here only applies to the disc-integrated flux.
Looking towards the limb, the contribution of metal opacity becomes larger
and the hydrogen opacity becomes smaller.  If the metal photoionisation
cross-sections are underestimated then the overall continuous opacity scale
factor must increase towards the limb. 
An alternative approach to that adopted here is
to correct the metal photoionisation cross-sections directly
\citep[e.g.][]{2022A&A...657L..11K}; in this case the 
scale factor would not change drastically going from
disc-centre to the limb.

\subsection{Step 2: Line list and blend calibration}
\label{resultsblend}

Following \citet{2022A&A...657L..11K}, the VALD
database was used to generate a theoretical
line list to describe the \ion{Be}{II} $313.107\,\nm$ region.
This line list includes $74$ known blends plus the
\ion{Be}{II} line itself;
the blue-most line is
the strong \ion{Ti}{II} $313.080\,\nm$,
and the red-most line is
the \ion{Tm}{II} $313.126\,\nm$.
Repeating the entire analysis without any lines except
for the \ion{Be}{II} $313.107\,\nm$ and the calibrated 
hypothetical line discussed below changed the overall
result by just $+0.037\,\dex$.  Half of this value
was folded into the uncertainty
on the overall result (\sect{resultsabundance}).

Apart from the hypothetical line discussed below,
only one empirical and two theoretical modifications
were made to this initial line list.
The empirical modification was for the 
position of the \ion{Cr}{I} line that blends the 
red wing of the \ion{Be}{II} $313.107\,\nm$.
It was shifted from its nominal laboratory wavelength of $313.1216\,\nm$ 
to the value given in
\citet{1975A&A....42...37C}, namely $313.1207\,\nm$.
The theoretical modifications concern
the parameters for broadening due 
to elastic collisions with neutral hydrogen
(usually referred to as van der Waals broadening)
for the \ion{Ti}{II} $313.080\,\nm$ 
and \ion{Cr}{I} $313.121\,\nm$ lines
that perturb the wings of the 
\ion{Be}{II} $313.107\,\nm$ feature.
In this work, these were calculated using ABO theory.
The \ion{Ti}{II} line broadening was calculated
with the theory extended to singly ionised 
species as in \citet{1998MNRAS.300..863B},
assuming the Uns\"{o}ld approximation value for
the parameter $E_{p}=-4/9$ atomic units, 
which should be expected to be a reasonable approximation for most lines 
(see Figure 2 of \citealt{2005A&A...435..373B}).
The \ion{Cr}{I} line broadening
was calculated with the theory as presented 
in \citet{1995MNRAS.276..859A}.
The recommended numbers are
$\{\sigma,\alpha\}=\{217\,\mathrm{a_{0}^2},0.213\}$
and $\{415\,\mathrm{a_{0}^2},0.262\}$ respectively,
where $\sigma$ is the broadening cross-section
at reference velocity $\varv=10^{4}\,\mathrm{m\,s^{-1}}$,
and $\alpha$ is the dimensionless exponent of the velocity scale factor
$\varv^{-\alpha}$.
These parameters were found to improve the quality of
the fits compared to the LFU theory with no enhancement factor.

The most influential adjustment to the line list was 
the introduction of a hypothetical line
to explain the observed width and asymmetry of the 
\ion{Be}{II} $313.107\,\nm$ feature.
By analysing the feature in dwarfs and giants,
\citet{2018ApJ...865....8C} argued that this blend could be caused
by a line of an ionised species with a low excitation energy 
for the lower level of the transition.
\citet{1975A&A....42...37C} arrived at a similar conclusion
on analysing the solar centre-to-limb variation.
If true, this is fortuitous in the sense that 
generally weak, low-excitation lines of majority ionised species,
typically show small departures from LTE in solar-metallicity
dwarfs (e.g.~for iron see 
\citealt{2022A&A...668A..68A}; for titanium
see \citealt{2022A&A...668A.103M,2024A&A...687A...5M}).
The authors adopted \ion{Ti}{II}
with $E_{\mathrm{low}}=0.01\,\mathrm{eV}$;
and this was adopted in this work.
A first guess of the laboratory wavelength was taken 
from \citet{1975A&A....42...37C}, namely $313.1019\,\nm$.

The disc-centre ($\mu=1.0$) and limb ($\mu=0.2$) 
data presented in \citet{1975A&A....42...37C}
were used to calibrate the wavelength and oscillator strength of 
the hypothetical \ion{Ti}{II} line.  These observational data were
extracted via a digital conversion from their Figures 5 and 6 respectively.  
Their continuum placement was adopted without modification,
but the wavelengths were shifted to the blue
by a velocity of $633\,\mathrm{m\,s^{-1}}$
to correct for the solar gravitational redshift. As discussed in
\citet{1975A&A....42...37C}, the asymmetry in the \ion{Be}{II} $313.107\,\nm$
feature can be clearly seen in the line intensities, which are not broadened by
solar rotation.  This facilitates a more robust calibration of the blend,
especially when combined with 3D non-LTE models, which incorporate other line
asymmetries due to solar surface convection and for which microturbulence and
macroturbulence do not need to be calibrated.  
As we mentioned at the start of the present Section, the nominal
resolving power of the spectrograph is $R=3\times10^{5}$.  
In the analysis the theoretical spectra were broadened using
a sinc$^{2}$ kernel of full width at half maximum $\lambda/R$,
and this led to an improved fit for the line wings.

The calibration proceeded as follows.
The \citet{1975A&A....42...37C} observations at 
disc-centre ($\mu=1.0$) and at the limb ($\mu=0.2$)
were fit simultaneously.  
The fit initially had five free parameters.  Three of them, namely
the beryllium abundance, and the oscillator strength and the wavelength of 
the hypothetical \ion{Ti}{II} line,
were constrained to be the same at both pointings.
The other two free parameters were the
missing continuous opacity factors, at disc-centre
and at the limb.  These were allowed
to vary between the two pointings
because the missing continuous opacity has
a centre-to-limb variation itself: there is an increasing
contribution of metal opacity towards the limb
as we discussed in \sect{resultsopacity}.
We note that the initial best-fitting beryllium abundance
from this analysis of the disc-centre and limb 
alone was $\lgeps{Be}=1.16$.
Afterwards, the calibration was repeated but with the
beryllium abundance constrained to the
best-fitting value following the analysis
of the disc-integrated flux
(\sect{resultsabundance}); this was iterated
until convergence was achieved.
Folding in the disc-integrated flux,
for which we directly calibrated the missing opacity
(\sect{resultsopacity}), raises the 
inferred beryllium abundance to our final
value of $\lgeps{Be}=1.21$ 
as we discuss in \sect{resultsabundance}.

We show the fits in \fig{fig:blend}, along with the fitting mask.
Apart from the hypothetical blend,
the blue wing of the \ion{Be}{II} $313.107\,\nm$ line is perturbed by the 
strong \ion{Ti}{II} $313.080\,\nm$ line 
together with a weaker OH line at $313.093\,\nm$.
At disc-centre, these features are reproduced well by the models.
Consequently, expanding the fitting region by $0.005\,\nm$ 
to include the far blue wing of the feature
had a negligible change on
the calibrated oscillator strength of the blend.
The agreement for the red wing is less good, but remains
satisfactory and is an improvement over what was found by
\citet{1975A&A....42...37C}.  There is significant overlap with 
the weak \ion{Zr}{I} $313.111\,\nm$ line,
and the wing is more strongly perturbed by
the combination
of the \ion{Cr}{I} $313.121\,\nm$, \ion{Fe}{I} $313.124\,\nm$, and 
\ion{Tm}{II} $313.126\,\nm$ lines
The first of these, the \ion{Zr}{I} 
$313.111\,\nm$ line, was discussed
by \citet{2018ApJ...865....8C} who suggested to reduce 
its oscillator strength.  
Doing so would only slightly decrease the beryllium
abundance inferred in this study. 
At the far limb, 
the blue wing is reproduced much better than the red wing.
This could reflect the growing importance of non-LTE
effects towards the limb, and how
the \ion{Ti}{II} and OH lines
on the blue wing form relatively close to LTE.
The minority species
\ion{Cr}{I} and \ion{Fe}{I} on the red wing 
are expected to suffer from overionisation,
and these lines may therefore appear weaker if modelled
in 3D non-LTE.

The main limitation of this calibration
arises from the treatment of the missing continuous opacities
at disc-centre and at the limb.
These were allowed to vary freely and separately, and this could
compensate for residual deficiencies in the 3D non-LTE models
of the centre-to-limb variation.
While 3D non-LTE models consistently
outperform 3D LTE, 1D non-LTE, and especially 1D LTE models
\citep[e.g.][]{2017MNRAS.468.4311L,2024A&A...683A.242C},
it is possible that improved physics are 
needed to obtain more accurate results,
such as magnetic fields
\citep{2023A&A...679A..65L,2024NatAs.tmp...73K}.
Forcing the beryllium abundance and the blend
parameters to be the same at disc-centre and the limb
helps mitigate the possible effects of this and
helps to get a more reliable calibration for the blend.
As a limiting test, the analysis was repeated 
forcing the continuous opacity scale factors
at disc-centre and limb to be that calibrated
for the disc-integrated flux, $1.17$ (\sect{resultsopacity}).
The fits in that case were poor at the limb; nevertheless,
the overall solar beryllium abundance changed by
only $+0.049\,\dex$. Half of this value was folded into the uncertainty
on the overall result (\sect{resultsabundance}).

We show the final line list in
\tab{tab:linelist}. The provided line list includes
the empirical modifications and calibrations discussed above. 

\subsection{Step 3: The solar beryllium abundance and its uncertainty}
\label{resultsabundance}

With the calibration for the 
missing continuous opacity in the disc-integrated flux
(a factor of $1.17$ applied to the continuous
opacity in the spectrum synthesis with \scate{};
see \sect{resultsopacity}) and
parameters for the blending line
calibrated on the centre-to-limb variation
(tuned wavelength and oscillator strength; see
\sect{resultsblend}), the solar beryllium
abundance was derived by fitting
the high resolution solar flux atlas from \citet{2005MSAIS...8..189K},
which is a re-reduction of the atlas of
\citet{1984sfat.book.....K}.\footnote{\url{http://kurucz.harvard.edu/sun/fluxatlas2005/solarfluxintwl.asc}.}
The continuum was placed by comparison with the normalised 
centre and limb spectra from \citet{1975A&A....42...37C}.
Rotational broadening was taken into account in the theoretical spectra
following Section 2 of \citet{1990A&A...228..203D},
adopting $\vsini=2.0\,\kms$ \citep{2016A&A...592A.156D}.
Like with the intensities (\sect{resultsblend}),
these rotationally broadened disc-integrated fluxes were
then broadened further using
a sinc$^{2}$ kernel of full width at half maximum $\lambda/R$,
with $R=3\times10^{5}$.

We show the fit in \fig{fig:flux}.
The best-fitting value for the solar beryllium abundance
is $1.21\pm0.05\,\dex$.
This value is significantly lower than the
value of \citet{2021A&A...653A.141A}, $1.38\pm0.09$, 
in large part because the original analysis in
\citet{2004A&A...417..769A} neglected
the impact of the hypothetical blend as we discussed in \sect{introduction}.
It is also lower than values obtained by other
recent 1D LTE and 1D non-LTE studies
\citep{2018ApJ...865....8C,2022A&A...657L..11K},
of around $1.30$ to $1.35\,\dex$,
which we attribute primarily to differences in the calibration strategy:
our value is closer to the value of $1.15\,\dex$ found by
\citet{1975A&A....42...37C} who also calibrated the blend on high-resolution
observations of the disc-centre and limb intensities, albeit in 1D non-LTE and
with a small nine-level model atom.

The uncertainty of $0.05\,\dex$ was estimated by first running 
$1\,000$ further fits perturbing different ingredients
by drawing from independent Gaussian distributions of
different standard deviations $\sigma$
and taking the standard deviation of the 
$1\,000$ resulting beryllium abundances.
The ingredients that varied were 
the scale factor for the continuous opacity
($\sigma=0.05$; as we discussed
in \sect{resultsopacity}), as well as a scale factor
for the continuum placement ($\sigma=0.03$).
The resulting uncertainty was $0.036\,\dex$. Secondly, the 
uncertainty caused by other blending lines was estimated 
as half the difference with the abundance obtained by repeating
the entire analysis without any blends apart from the calibrated
one (as we discussed in \sect{resultsblend}); this gave $0.018\,\dex$.
Lastly, the effect of the uncertainty in the blend calibration
was estimated as half the difference with the result
obtained when the continuous opacity scale factors
at disc-centre and limb were fixed at $1.17$ (as we discussed in
\sect{resultsblend});
this gave $0.025\,\dex$.
The final uncertainty estimate was obtained by adding these
three numbers in quadrature.

\section{The depletion of beryllium in comparison with solar model predictions}
\label{discussion}

\begin{figure}
    \begin{center}
        \includegraphics[scale=0.425]{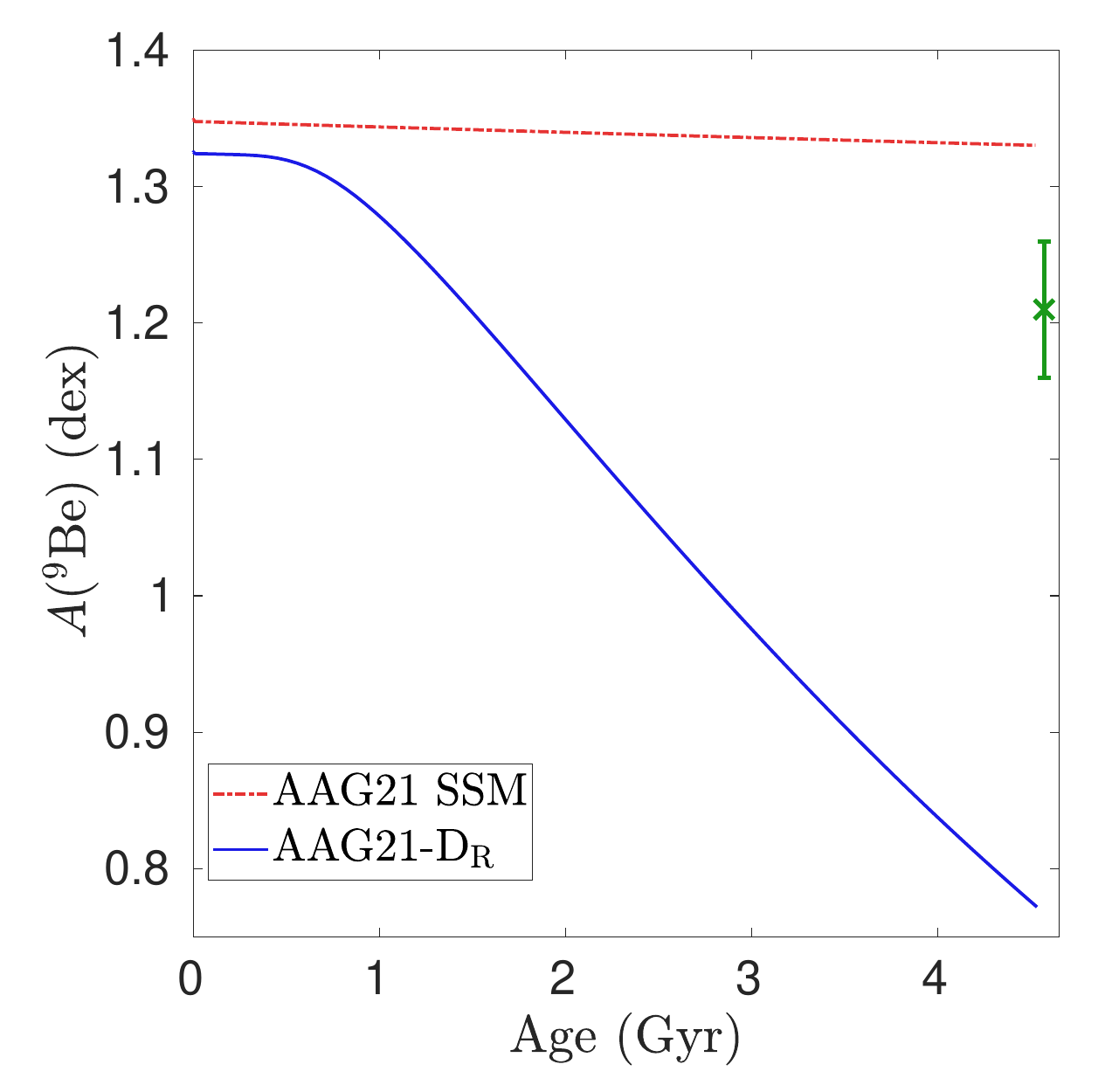}
        \caption{Predicted evolution of the surface
        beryllium abundance with different theoretical models.
        Red shows a standard solar model (SSM);
        blue shows a model that includes the
        effects of angular momentum transport (D$_{\mathrm{R}}$). 
        The point at $4.6\,\mathrm{Gyr}$ shows
        the current value of the surface beryllium abundance
        that was determined in \sect{resultsabundance}.}
    \label{fig:modelbe}
    \end{center}
\end{figure} 

The extra depletion of beryllium in the solar surface
(on top of any effects of atomic diffusion)
is estimated to be $-0.11\pm0.06\,\dex$, or $22\pm11\%$,
when taking $\lgeps{Be}=1.21\pm0.05$ that we found in
\sect{results}, and 
$\lgeps{Be}_{\text{init}}=1.32\pm0.04$ as we discussed 
in \sect{introduction}.
In this section we compare these results to predictions
from a standard solar model (SSM) as well as a solar
model that includes the effects of angular momentum transport
(D$_{\mathrm{R}}$),
both computed using the Li\`ege stellar evolution code (\cles{}; 
\citealt{2008Ap&SS.316...83S}).

The D$_{\mathrm{R}}$ model includes the combined effects of hydrodynamical and
magnetic instabilities following an asymptotic formalism introduced in
\citet{2024A&A...686A.108B}.
This analytical expression allows to capture the full
effects of the combined meridional circulation, shear-induced turbulence and
magnetic Tayler instability 
\citep{2002A&A...381..923S} on the depletion of light
elements. In \citet{2022NatAs...6..788E}, the combination of these physical
mechanisms was used to explain simultaneously the solar rotation profile, the
observed photospheric lithium depletion 
and an increased helium mass fraction in the solar convective zone. 
Compared to the model of \citet{2022NatAs...6..788E}
based on the Geneva stellar evolution code
(\genec{}; \citealt{2008Ap&SS.316...43E}),
the \cles{} model used here predicts a larger amount of beryllium depletion.
This is caused by a combination of factors, including
differences in the nuclear reaction network between the two codes,
as well as different reference solar abundances ---
rather than \citet{2009ARA&A..47..481A}, the \cles{} model used here is 
based on the compilation of \citet{2021A&A...653A.141A},
which is overall more metal-rich and also
has a $0.09\,\dex$ lower
lithium abundance (via \citealt{2021MNRAS.500.2159W}).

We illustrate the predicted evolution of the surface abundance
of beryllium in \fig{fig:modelbe}.  
The initial chemical compositions of the two models
do not match exactly, as they are free parameters in the models
and thus are sensitive to the different input physics.
As such, the relevant quantity to inspect in these
plots is the relative difference between the
initial and final beryllium abundances
(i.e.~the beryllium depletion itself).
It turns out that neither the SSM nor the D$_{\mathrm{R}}$ model
can reproduce the beryllium depletion reported in this work:
the former predicts negligible depletion,
while the latter predicts a depletion of more than
$0.5\,\dex$ (roughly a factor of three).

One possible explanation is that the initial solar
beryllium abundance is in error.
The adopted value of $\lgeps{Be}_{\text{init}}=1.32\pm0.04$ via
\citet{2021SSRv..217...44L} is based on the Orgueil
meteorite.
The analysis of the Ivuna meteorite by \citet{2020GeCoA.268...73K}
gives a larger beryllium abundance:
$\lgeps{Be}_{\text{init}}=1.38\pm0.04$,
using silicon, magnesium, and iron to convert
the meteoritic abundances onto the solar scale
(see \sect{introduction}).
As discussed therein, Ivuna is possibly less affected by 
terrestrial modification. However,
The two meteorites typically agree well,
with a mean difference in 
$\mathrm{X/Si}$ ratios of $-0.02\pm0.03\,\dex$ for 52 elements
in common, and the discrepancy could be due to the heterogeneous
nature of CI chondrites and the small sample size
of the Ivuna meteorite.
In any case, using the result from the Ivuna meteorite
the beryllium depletion in the Sun would be
$0.17\pm0.04$, which is still far less than that
predicted by the D$_{\mathrm{R}}$ model.
We also note that the effects of atomic diffusion
are neglected in the conversion from the meteoritic abundance scale to the solar
abundance scale in \sect{introduction}. This may amount to a $0.03\,\dex{}$ 
to $0.06\,\dex{}$ larger value of
$\lgeps{Be}_{\text{init}}$
(see Section 5 of \citealt{2021A&A...653A.141A}).
Nevertheless, the corresponding
depletion values would still be far too low to be explained by the 
D$_{\mathrm{R}}$ model.

The failure of the D$_{\mathrm{R}}$ model is not
too surprising as the so-called
Tayler-Spruit instability has proven unable to reproduce asteroseismic
constraints on the internal rotation of young solar-like subgiants
\citep{2020A&A...641A.117D}.
Further investigations will be required to determine
whether a combination of other processes such as overshooting or opacity
increases at the base of the solar convective zone, as investigated in
\citet{2017ARep...61..901A},
\citet{2019A&A...621A..33B}, and \citet{2022A&A...667L...2K},
will be able to reconcile these new solar models with observations.

Detailed comparisons between various evolution codes 
(e.g.~Deal, in prep.) will shed new light on the
predicted light element depletions in the context of the solar modelling
problem.  Nevertheless, the observed depletion of beryllium in the Sun
advocates for
some form of turbulent mixing at the base of the solar convective zone acting
during the main-sequence, in disagreement with the prescriptions used in
standard solar models that only consider microscopic diffusion. After having
calibrated the transport coefficients on the solar depletion, it might also be
interesting to apply them to solar twins, such as the 16 Cyg binary system
\citep{2020A&A...644A..37F,2022A&A...661A.143B} for which
\citet{2015A&A...584A.105D} used the light element
depletion as a tracer of planetary accretion.

\section{Conclusions}
\label{conclusion}

We have presented 3D non-LTE calculations for beryllium in the solar atmosphere,
and have used the models to determine the present-day 
abundance of beryllium in the solar surface. The analysis
presented here combines and builds upon ideas presented separately in the
literature: it uses the observed solar irradiance to calibrate the missing
opacity \citep{2022A&A...657L..11K}, the centre-to-limb variation to calibrate
the blend \citep{1975A&A....42...37C}, which in turn is assumed to be a
low-excitation line of an ionised majority species based on the
analyses of \citet{1975A&A....42...37C} and
\citet{2018ApJ...865....8C}.

We found $\lgeps{Be}=1.21\pm0.05$ for the present-day surface abundance.  The
initial abundance was taken to be
$\lgeps{Be}_{\text{init}}=1.32\pm0.04$ based
on the CI chondrite Orgueil and converted to the solar scale using
silicon, magnesium, and iron
and neglecting the possible impact of atomic diffusion;
although we note that a recent analysis of the CI chondrite Ivuna 
could indicate an even larger initial abundance.
With this initial value we found that beryllium
has been depleted by 
an extra $0.11\pm0.06\,\dex$, or $22\pm11\%$,
on top of any effects of atomic diffusion. This
This is in tension with
standard solar models, which predict negligible depletion,
as well as contemporary solar models
with calibrated extra mixing \citep[e.g.][]{2022NatAs...6..788E},
which predict excessive depletion;
the non-Standard 
model used here predicts a depletion in excess of $0.5\,\dex$,
for example.

To even better pin down the solar beryllium abundance in the near future,
tighter constraints on the line list would be welcome.
This could be tackled empirically,
for example by studying other stars of different stellar
parameters like in \citet{2018ApJ...865....8C},
but using 3D non-LTE models to avoid
the microturbulence or macroturbulence fudge parameters
that could hamper calibrations.
Furthermore, improved 3D non-LTE modelling of the centre-to-limb variation,
for example using models including magnetic fields
\citep{2023A&A...679A..65L,2024NatAs.tmp...73K},
would help ensure that the calibration presented here
is indeed reliable.  On longer timescales, a better first-principles
description of the
continuous opacity in the UV as well as a concrete identification
of the $313.102\,\nm$ blend are sorely needed.
In particular, theoretical atomic structure calculations
could perhaps shed light on the identity of the blending feature,
for instance to confirm or rule out 
the \ion{Mn}{I} $313.104\,\nm$ line as a contender.
In the meantime,
we make the 3D LTE and 3D non-LTE spectra
used in this work publicly available 
(online Table 2) so that others may
use them to explore different calibration strategies.

\begin{acknowledgements}
    The authors wish to thank S.~Korotin for providing a
    constructive referee report.
    AMA acknowledges support from the Swedish Research Council (VR 2020-03940)
    and from the Crafoord Foundation via the
    Royal Swedish Academy of Sciences (CR 2024-0015).  GB is funded by
    an Fonds National de La Recherche Scientifique (FNRS) postdoctoral
    fellowship.  
    YZ gratefully acknowledges support from the Elaine P.~Snowden 
    Fellowship.
    PSB acknowledges support from the Swedish Research Council
    (VR 2020-03404).
    This research was supported by computational resources provided
    by the Australian Government through the National Computational
    Infrastructure (NCI) under the National Computational Merit Allocation
    Scheme and the ANU Merit Allocation Scheme (project y89).  Some of the
    computations were also enabled by resources provided by the National
    Academic Infrastructure for Supercomputing in Sweden (NAISS), partially
    funded by the Swedish Research Council through grant agreement no.
    2022-06725, at the PDC Center for High Performance Computing, KTH Royal
    Institute of Technology (project number PDC-BUS-2022-4).
\end{acknowledgements}

\bibliographystyle{aa_url} 
\bibliography{bibl.bib}

\end{document}